\documentclass[cup6b]{cupbook}
\usepackage{natbib}
\usepackage{natpatch}
\usepackage{epsf}
\usepackage{graphicx}

\newcommand{\beq}{\begin{equation}}
\newcommand{\eeq}{\end{equation}}
\newcommand{\bea}{\begin{eqnarray}}
\newcommand{\ena}{\end{eqnarray}}

\newcommand{\xisigmav}{\xi^2\langle \sigma_{\rm ann} v \rangle}

\newcommand{\pbar}{\mbox{$\bar{\rm p}$}}

\newcommand{\nbar}{\mbox{$\bar{\rm n}$}}
\newcommand{\dbar}{\mbox{$\bar{\rm D}$}}

\newcommand{\lD}{\mbox{$\lambda_{\rm D}$}}

\newcommand\lsim{\mathrel{\rlap{\lower4pt\hbox{\hskip1pt$\sim$}}
    \raise1pt\hbox{$<$}}}
\newcommand\gsim{\mathrel{\rlap{\lower4pt\hbox{\hskip1pt$\sim$}}
    \raise1pt\hbox{$>$}}}
\def\lsim{\mathrel{\raise.3ex\hbox{$<$\kern-.75em\lower1ex\hbox{$\sim$}}}} 
\def\gsim{\mathrel{\raise.3ex\hbox{$>$\kern-.75em\lower1ex\hbox{$\sim$}}}}

\newcommand{\ie}{{\it i.e.}}

\newcommand{\svave}{\langle\sigma_{\rm ann} v\rangle}

\begin{document}


\author[P. Salati, F. Donato and N. Fornengo]{Pierre Salati$^a$, Fiorenza Donato$^b$, Nicolao Fornengo$^b$ 
\\ $^a$ Laboratoire d'Annecy--le--Vieux de Physique Th\'eorique ({\sc LAPTH}), 
  CNRS--IN2P3 \\ Universit\'e de Savoie, Chemin de Bellevue, 
  74000 Annecy--le--Vieux Cedex 09, France \\ 
  $^b$ Dipartimento di Fisica Teorica, Universit\`a di Torino 
\\ and INFN--Sezione di Torino,
  Via P. Giuria 1, 10122 Torino, Italy}

\chapter{Indirect Dark Matter Detection with Cosmic Antimatter}

\section{Production of antimatter in the galaxy}
The indirect detection of particle dark matter (DM) is based on the search for anomalous
components in cosmic rays (CRs) due to the annihilation of DM pairs in the galactic halo, 
on the top of the standard astrophysical production. 
These additional exotic components are
potentially detectable at Earth as spectral distortions
for the various cosmic radiations:
\beq
\chi + \chi \to q \bar{q} , W^{+} W^{-} , \ldots \to
\bar{p} , \bar{D} , e^{+} \, \gamma \, \& \, \nu's \;\; .
\eeq
Detection of the DM annihilation products has motivated the spectacular
development of several new experimental techniques. They range from detectors on
ballons or in space for the study of antimatter and gamma--rays, to large area cosmic--ray
and gamma--ray detecors on the ground to neutrino
telescopes underground for the study of the neutrino component. In the following, we will discuss
in detail the antimatter component of DM indirect searches, namely antiprotons, antideuterons,
and positrons.

\section{Propagation of antinuclei in the Galaxy}

\begin{figure*}[t]
\begin{center}
\noindent
\includegraphics[width=0.95\textwidth]{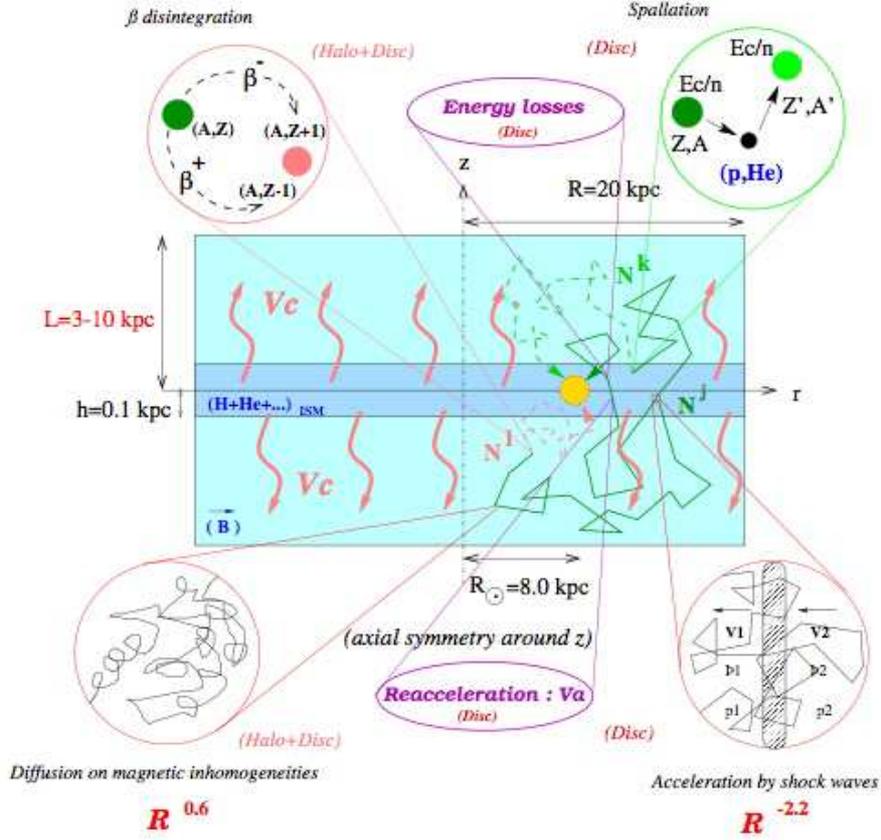}
\end{center}
\vskip -0.5cm
\caption{
Schematic edge--on view of the Milky Way diffusive halo (DH) as seen by a cosmic ray
physicist. The stellar and gaseous disc is sandwiched between two thick layers which
contain turbulent magnetic fields.
After having been accelerated by SN driven shock waves or produced by DM species
annihilating in the galactic halo, cosmic rays diffuse on magnetic inhomogeneities
and are wiped away by a galactic wind with velocity $V_{C}$. They can lose energy
and are also mildly subject to diffusive reacceleration. The former process is by
far the dominant one in the case of electrons and positrons.
This diagram has been borrowed from the review~\cite{Maurin:2002ua}.
}
\label{CR_camembert}
\end{figure*}
Whatever the mechanism responsible for their production, charged cosmic
rays subsequently propagate through the galactic magnetic field and are
deflected by its irregularities~: the Alfv\'en waves. In the regime where
the magnetic turbulence is strong -- which is the case for the Milky Way
-- Monte Carlo simulations \cite{Casse:2001be}
 indicate that it is similar to space diffusion with a coefficient:
\beq
K(E) \, = \, K_{0} \; \beta \;
\left( {\mathcal R}/{\rm 1 \; GV} \right)^{\delta} \;\; ,
\label{space_diffusion_coefficient}
\eeq
which increases as a power law with the rigidity ${\mathcal R} = {pc}/{Ze}$ of the
particle.
In addition, because the scattering--centers drift inside the Milky Way with
a velocity $V_{a} \sim$ 20 to 100 km s$^{-1}$, a second--order Fermi--mechanism
is  responsible for some mild diffusive reacceleration. Its coefficient $K_{EE}$
depends on the particle velocity $\beta$ and total energy $E$ and is
related to the space diffusion coefficient $K(E)$ through:
\beq
K_{EE} \, = \,
{\displaystyle \frac{2}{9}} \; V_{a}^{2} \;
{\displaystyle \frac{E^{2} \beta^{4}}{K(E)}} \;\; .
\eeq
%
%
Finally, galactic convection wipes cosmic rays away from the disc with a velocity
$V_{C} \sim$ 5 to 15 km s$^{-1}$.
We can assume steady state for the various populations of particles and write the master
equation for the space and energy distribution function $\psi = dn / dE$ as
\cite{1990acr..book.....B}:
\beq
\partial_{z} \left( V_{C} \, \psi \right)  \, - \, K  \, \Delta \psi \, + \,
\partial_{E} \left\{
b^{\rm loss}(E) \, \psi  \, - \, K_{EE}(E) \, \partial_{E} \psi \right\}
\, = \, q \left( {\mathbf x} , E \right) \;.
\label{master_equation}
\eeq
This equation applies to any charged species -- nuclei, protons, antiprotons or positrons
-- as long as the rates for production $q$ and energy loss $b^{\rm loss}(E)$
are properly accounted for.

The solution of the master equation \ref{master_equation}
 has been deeply investigated and several different techniques 
lead to very similar fluxes at the Earth
\cite{Strong:2007nh,Maurin:2002ua}. 
One possibility is a completely numerical solution. This is the way followed 
in the Galprop model \cite{1998ApJ...509..212S,Strong:2004de} which, thanks to a realistic gas distribution, 
can calculated gamma rays in addition to charged species\cite{Porter:2008ve}. 
A completely different approach of the CRs transport through the DH relies 
on the calculation of the Green function of Eq. \ref{master_equation}, which 
describes the probability for a CR that is produced at a location $\vec{x}$ 
with energy $E_S$ to be detected at the Earth with a degraded energy $E$. The
Green function is then integrated over the diffusive volume (DH) and the
energy range. This is {\it i.e.} the approach chosen to propagate positrons 
\cite{1998ApJ...493..694M,2008PhRvD..77f3527D,eplus_sec}. 

The Bessel expansion method, based on the cylindrical symmetry of the DH and 
on approximate values for the ISM (not relevant for charged CR propagation)
permits a 2D fully analytical model. Numerical solution is required only for the
diffusion in energy space. This is the model developed
in~\cite{Maurin:2001sj,Maurin:2002ua,Maurin:2002hw,Donato:2001eq} and 
detailed more extensively in the following of this chapter. 
Very similar results for the propagation of stable primary and secondary
nuclei have been obtained with the modified weighted slab
technique~\cite{2001ApJ...547..264J}.

According to this approach the region of the
Galaxy inside which cosmic rays diffuse -- the so--called diffusive halo (DH)
-- is pictured as a thick disc which matches the circular structure of the Milk Way
as shown in Fig.~\ref{CR_camembert}. The galactic disc of stars and gas,
where primary cosmic rays are accelerated, lies in the middle. It extends
radially 20 kpc from the center and has a half--thickness $h$ of 100 pc.
Confinement layers where cosmic rays are trapped by diffusion lie above and beneath
this thin disc of gas.
The intergalactic medium starts at the vertical boundaries $z = \pm L$ as well as
beyond a radius of $r = R \equiv 20$ kpc. The half--thickness $L$
of the diffusive halo is not known and reasonable values range from 1 to 15 kpc.
The diffusion coefficient $K$ is the same everywhere whereas the convective velocity
is exclusively vertical with component $V_{C}(z) = V_{C} \; {\rm sign}(z)$. This galactic
wind, which is produced by the bulk of disc stars like the Sun, drifts away from its
progenitors along the vertical directions.
The normalization coefficient $K_{0}$, the index $\delta$, the
galactic drift velocity $V_{C}$ and the Alfv\'en velocity $V_{a}$ may be
determined by studying the boron--to--carbon ratio (B/C) which
is quite sensitive to cosmic ray transport and which may be used efficiently as a constraint.

The Bessel expansion method takes advantage of the axial
symmetry of the DH and enforces a vanishing cosmic ray flux at
a distance $R = 20$ kpc from the rotation axis of the Galaxy. This condition
is actually implemented naturally by the following series expansion for $\psi$:
\beq
\psi \left( r , z , E \right) \, = \,
{\displaystyle \sum_{i=1}^{+ \infty}} \; P_{i} \left( z , E \right) \,
J_{0} \left( \alpha_{i} \, r / R \right) \;\; .
\label{bessel_psi}
\eeq
The Bessel function of zeroth order $J_{0}$ vanishes at the points $\alpha_{i}$.
The radial dependence of $\psi$ is now taken into account by the set of its Bessel
transforms $P_{i}(z,E)$. The source term $q$ may also be Bessel expanded into
the corresponding functions $Q_{i}(z,E)$ so that the master
equation~(\ref{master_equation}) becomes
\begin{eqnarray}
\partial_{z} \left( V_{C} \, P_{i} \right) & - &
K \, \partial_{z}^{2} P_{i} \, + \,
K \, \left\{ {\displaystyle \frac{\alpha_{i}}{R}} \right\}^{2} \! P_{i} \, + \,
\label{master_2} \\
& + &
{2  \, h \, \delta(z)} \,
\partial_{E} \! \left\{ b^{\rm loss}(E) \, P_{i} \, - \,
K_{EE}(E) \, \partial_{E} P_{i} \right\}
\, = \, Q_{i} \left( z , E \right) \;\; .
\nonumber
\end{eqnarray}
Here, energy loss and diffusive reacceleration are confined inside the galactic
disc -- which is considered infinitely thin, hence the presence of an effective
term $2 \, h \, \delta(z)$.
The form of the source terms $Q_{i}(z,E)$ which appear in equation~(\ref{master_2})
depends on the nature of the cosmic ray particle.

\section{Antiprotons in Cosmic Rays}
\label{pbar}

In the case of antiprotons, the following mechanisms can in principle
contribute to the source term in the transport equation \ref{master_equation}:

\begin{itemize}
\item
The spallation of high--energy primary nuclei impinging on the atoms 
of the interstellar medium inside the galactic disc produces 
secondary antiprotons.
\item
The annihilation of DM candidate particles throughout the Milky Way halo
generates primary antiprotons. Notice
that WIMP annihilations take place all over the diffusive halo.
\item
Tertiary antiprotons result from the inelastic and non--annihilating interactions 
with a nucleon at rest. The energy transfer may be sufficient 
to excite it as a $\Delta$ resonance. This mechanism redistributes antiprotons
toward lower energies and flattens their spectrum  \cite{1999ApJ...526..215B}.
This yields the source term:
\begin{eqnarray}
q_{\bar{\rm p}}^{\rm ter}(r , E_{\bar{\rm p}}) & = &
{\displaystyle \int_{E_{\bar{\rm p}}}^{+ \infty}} \,\,
{\displaystyle
\frac{d \sigma_{\rm \bar{p} \, H \to \bar{p} \, X}}{dE_{\bar{\rm p}}}}
(E'_{\bar{\rm p}} \to E_{\bar{\rm p}}) \; n_{\rm H} \; \beta'_{\bar{\rm p}} \;
\psi_{\bar{\rm p}}(r , E'_{\bar{\rm p}}) \; dE'_{\bar{\rm p}} \nonumber \\
& - & \;\;
\sigma_{\rm \bar{p} \, H \to \bar{p} \, X}(E_{\bar{\rm p}}) \; n_{\rm H} \;
\beta_{\bar{\rm p}} \; \psi_{\bar{\rm p}}(r , E_{\bar{\rm p}}) \;\; ,
\label{tertiary}
\end{eqnarray}
where the inelastic and non--annihilating differential cross section
in this expression can be approximated by:
\begin{equation}
{\displaystyle
\frac{d \sigma_{\rm \bar{p} \, H \to \bar{p} \, X}}{dE_{\bar{\rm p}}}}
\, = \, {\displaystyle
\frac{\sigma_{\rm \bar{p} \, H \to \bar{p} \, X}}{T'_{\bar{\rm p}}}} \;\; .
\end{equation}
The initial antiproton kinetic energy is denoted by $T'_{\bar{\rm p}}$.
In order to take into account elastic scatterings on helium, one simply has
to replace the hydrogen density by $n_{\rm H} \, + \, 4^{2/3} \, n_{\rm He}$.
\item
Antiprotons may also annihilate on interstellar H and He. This leads to
a negative source term $- \, \Gamma_{\bar{\rm p}}^{\rm ann} \, \psi$, where
the annihilation rate $\Gamma_{\bar{\rm p}}^{\rm ann}$ is defined as
\beq
\Gamma_{\bar{\rm p}}^{\rm ann} \, = \,
\sigma_{\bar{\rm p} \, {\rm H}}^{\rm ann}  \, \beta_{\bar{\rm p}} \, n_{\rm H} \, + \,
\sigma_{\bar{\rm p} \, {\rm He}}^{\rm ann} \, \beta_{\bar{\rm p}} \, n_{\rm He} \;\; .
\eeq
The annihilation cross section $\sigma_{\bar{\rm p} \, {\rm H}}^{\rm ann}$
can be borrowed from \cite{tan_ng_1982,1983JPhG....9..227T} and multiplied by a
factor of $4^{2/3} \sim 2.5$, taking into account the higher geometric
cross section, to get $\sigma_{\bar{\rm p} \, {\rm He}}^{\rm ann}$.
The average hydrogen $n_{\rm H}$ and helium $n_{\rm He}$ densities in
the galactic disc are respectively set equal to $0.9$ and $0.1$ cm$^{-3}$.
\end{itemize}

\subsection{Secondary antiprotons}
\label{pbar_sec}

Secondary antiprotons are produced by CR rays--spallation on the
interstellar medium. They represent the background when searching for small contributions 
coming from exotic sources, like signals from DM annihilation.

The rate for the production of secondary antiprotons takes the following form:
\beq
q_{\bar{\rm p}}^{\rm sec}(r , E_{\bar{\rm p}}) \, = \,
{\displaystyle \int_{E^{0}_{\rm p,\alpha}}^{+ \infty}} \,\, n_{\rm H,\alpha} \times
\beta_{\rm p,\alpha} \; \psi_{\rm p,\alpha}(r , E_{\rm p,\alpha}) \times dE_{\rm p,\alpha} \times
{\displaystyle \frac{d \sigma}{dE_{\bar{\rm p}}}}(E_{\rm p,\alpha} \to E_{\bar{\rm p}})
\;\; ,
\label{source_sec_pbar}
\eeq
for the interactions between cosmic ray protons and $\alpha$ particles,
and hydrogen and helium nuclei in the interstellar medium (ISM). 

Equation~(\ref{master_equation}) may be solved according to the method outlined
in Appendix B of \cite{2001ApJ...563..172D} for 
an antiproton source located only in the galactic disk, 
as is the case for secondary antiprotons. 

Some approximations may be at hand: in particular, setting the energy loss
rate $b^{\rm loss}$ and the energy diffusion coefficient $K_{EE}$ equal to
zero does not affect sizeably the solution of the diffusion equation. 

As for the the cosmic ray proton and helium fluxes, they can be borrowed
from~\cite{2001ApJ...563..172D}, where a fit  to high energy ($>$ 20 GeV/n)
data is proposed.  A more recent parameterization on IS fluxes has been derived
in~\cite{Donato:2008jk}.
 
\begin{figure*}[t]
\begin{center}
\includegraphics[width=0.48\columnwidth]{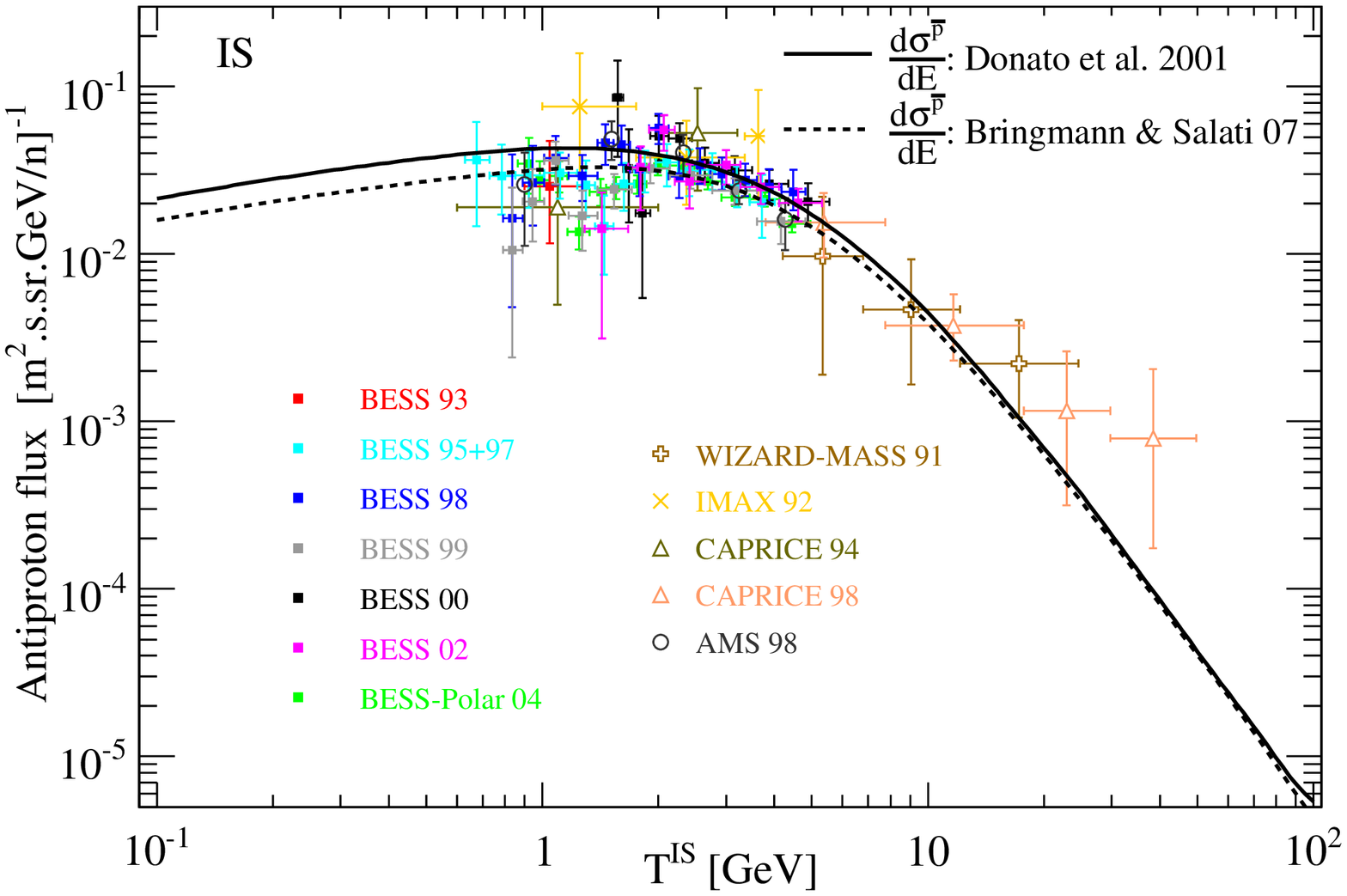}
\includegraphics[width=0.48\columnwidth]{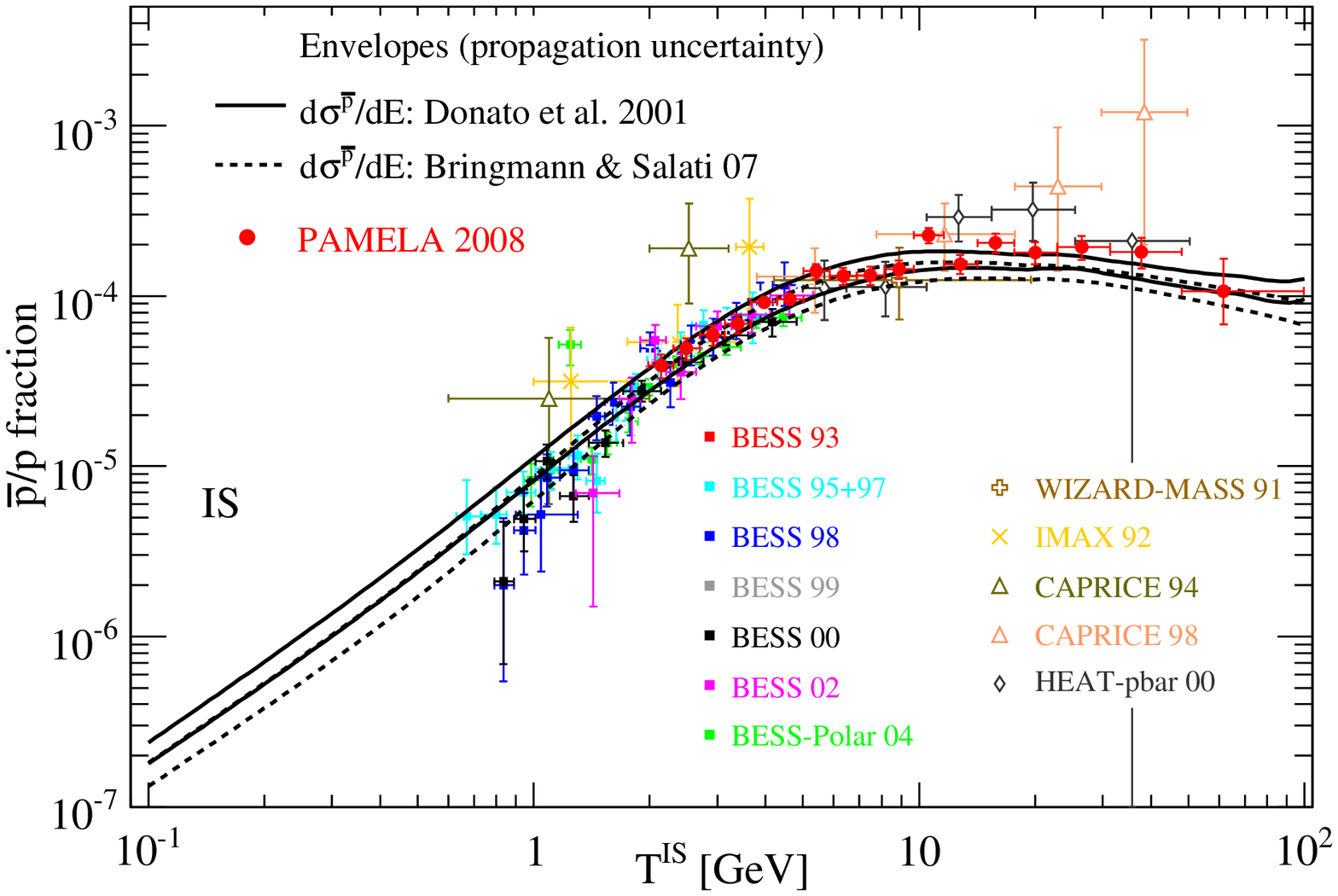}
\end{center}
\caption{Left panel: IS antiproton flux for the B/C best fit model and two
parameterizations of the production cross section.
Right panel: propagation uncertainty envelopes of the IS \pbar$/p$ ratio
for the same production cross sections as in the left panel.
All data are demodulated using the force-field approximation: AMS~98 \cite{2002PhR...366..331A},
IMAX~92 \cite{1996PhRvL..76.3057M}, CAPRICE~94 \cite{1997ApJ...487..415B},
WIZARD-MASS~91 \cite{1999ICRC....3...77B},  CAPRICE~98 \cite{2001ApJ...561..787B},
BESS93 \cite{1997ApJ...474..479M}, BESS~95+97 \cite{2000PhRvL..84.1078O},
BESS~98 \cite{2001APh....16..121M}, BESS~99 and 2000 \cite{2002PhRvL..88e1101A},
BESS~2002 \cite{2005ICRC....3...13H}, BESS Polar \cite{2008arXiv0805.1754A},
WIZARD-MASS~1~\cite{1996ApJ...467L..33H}, HEAT-$\bar{p}$ \cite{2001PhRvL..87A1101B},
and  PAMELA \cite{pamela_pbarp}.}
\label{fig:pbar_sec}
\end{figure*}

Once obtained the interstellar (IS) fluxes of antiprotons at
the Sun's position in the Galaxy, we have to further propagate them
inside the heliosphere, where the cosmic ray particles which
eventually reach the Earth are affected by the presence of the solar
wind.
We model the effect of solar modulation by adopting the force
field approximation of the full transport equation
\cite{1987A&A...184..119P}. In this model, the top--of--atmosphere (TOA) 
flux for a cosmic species $\Phi^{\rm TOA}$ is obtained as:
\begin{equation}
\frac{\Phi^{\rm TOA} (E^{\rm TOA})}{\Phi^{\rm IS} (E^{\rm IS})}
= 
\left( \frac{p^{\rm TOA}}{p^{\rm IS}} \right)^{2}\;
\end{equation}
where $E$ and $p$ denote the total energies and momenta of
interstellar and TOA antiprotons, which are related by the energy shift:
\begin{equation}
E^{\rm TOA} = E^{\rm IS} - \phi
\end{equation}
where the parameter $\phi$ is determined by fits on cosmic ray
data. A value $\phi=500$ MV is indicative of 
periods of minimal solar activity. 

The secondary IS \pbar\ flux is displayed in the left panel of Fig.~\ref{fig:pbar_sec}
along with the data demodulated according to the force-field prescription.
We either use the DTUNUC \cite{2001ApJ...563..172D} \pbar\ production cross sections
(solid line) or those discussed in Refs. \cite{Duperray:2005si,2007PhRvD..75h3006B} 
(dashed line).
The differences between the two curves illustrate the uncertainty related to the production
cross sections, as emphasized in \cite{2001ApJ...563..172D}, where a careful
and conservative analysis within the DTUNUC simulation settled a nuclear 
uncertainty of $\sim 25\%$ over the energy range $0.1-100$~GeV.
The conclusion is similar here, although the two sets of cross sections differ
mostly at low energy.
%
In the right panel,  along with the demodulated \pbar$/p$ data, we show the curves
bounding the propagation uncertainty on the \pbar\ calculation based either on the
DTUNUC \cite{2001ApJ...563..172D} \pbar\ production cross sections (solid lines)
or those borrowed from \cite{2007PhRvD..75h3006B} (dashed lines). The uncertainty
arising from propagation is comparable to the nuclear one \cite{2001ApJ...563..172D}.
The detailed
calculation of that secondary component \cite{2002A&A...394.1039M} has required the
determination of the propagation--diffusion parameters that are
consistent with the B/C data \cite{2001ApJ...555..585M}. By varying
those parameters over the entire range allowed by the cosmic--ray
nuclei measurements, the theoretical uncertainty on the antiproton
secondary flux has been found to be 9\% from 100 MeV to 1 GeV. It
reaches a maximum of 24\% at 10 GeV and decreases to 10\% at 100 GeV.

From Fig. \ref{fig:pbar_sec}, it is manifest that the secondary contribution alone explains
experimental data on the whole energetic range. 
It is not necessary to invoke an additional component
to the standard astrophysical one.

\begin{table*}[t]
\begin{center}
\vskip 0.15cm
\begin{tabular}{|l||c|c|c||c|c|}
\hline
\textbf{Halo model} & $\alpha$ & $\beta$ & $\gamma$ &
$\rho_{\mathrm{s}}$ [$10^{6} \; M_{\odot} \; \mathrm{kpc}^{-3}$] &
$r_{\mathrm{s}}$ [$\mathrm{kpc}$]\\
\hline
\hline
Cored isothermal~\cite{Bahcall:1980fb} & 2   & 2 & 0    & 7.90 & 4     \\
NFW 97~\cite{1997ApJ...490..493N}    & 1   & 3 & 1    & 5.38 & 21.75 \\
Moore 04~\cite{2004MNRAS.353..624D}  & 1   & 3 & 1.16 & 2.54 & 32.62 \\
\hline
\end{tabular}
\end{center}
\caption{
Parameters in equation~(\ref{eq:DM_profile_a}) for different halo models.
 The scale radius $r_{\mathrm{s}}$ and density
$\rho_{\mathrm{s}}$ are strongly correlated with the virial mass of the
Galaxy \cite{2001ApJ...554..114E} and the values are borrowed from \cite{Fornengo:2004kj}
for the Milky Way. 
When the DM distribution is cuspy ($\gamma \geq 1$) the divergence at
the galactic center is smoothed according to the prescription of
\cite{2007PhRvD..75h3006B}.}
\label{tab_WIMP_halo_a}
\end{table*}

\subsection{Antiprotons from DM annihilation}
\label{pbar_prim}

\begin{table*}[t]
\begin{tabular}{|c||c|c|c|c|c|c|}
\hline
Case  & $\delta$ & $K_0$ [kpc$^2$/Myr] & $L$ [kpc] & $V_{C}$ [km/s] & $V_{a}$ [km/s] \\
\hline \hline
MIN  & 0.85 &  0.0016 & 1  & 13.5 &  22.4 \\
MED  & 0.70 &  0.0112 & 4  & 12   &  52.9 \\
MAX  & 0.46 &  0.0765 & 15 &  5   & 117.6 \\
\hline
\end{tabular}
\vskip -0.25cm
\caption{Typical combinations of diffusion parameters that are compatible with the B/C
analysis \cite{2001ApJ...555..585M}. As shown in \cite{2004PhRvD..69f3501D}, these propagation
models correspond respectively to minimal, medium and maximal primary antiproton
fluxes.}
\label{tab_prop}
\end{table*}

The antiproton signal from annihilating DM particles leads to a primary component
directly produced throughout the DH. The 
differential rate of production per unit volume and time is a function of
space coordinates and
antiproton kinetic energy $T_{\bar p}$. It is defined as:
\begin{equation}
q_{\bar p}^{\rm DM}(r,z,T_{\bar p}) =
\, \xi^2\svave \, g(T_{\bar p})
\left( \frac{ \rho_\chi (r,z)}{m_\chi} \right)^2 ,
\label{eq:source_pbar_prim}
\end{equation}
where $\svave$ denotes the average over the Galactic velocity
distribution function of the WIMP pair annihilation cross
section $\sigma_{\rm ann}$ multiplied by the relative velocity $v$.
For a relic CDM particle able to explain the observed amount of cosmological dark matter
\cite{Hinshaw:2008kr,Komatsu:2008hk,Dunkley:2008ie} its value
falls in the range $2 \div 3 \cdot 10^{-26}$ cm$^3$ s$^{-1}$
(unless special situations occurs, like dominant p--wave annihilation, dominant coannihilations
or modified cosmology). For any DM candidate, $\sigma_{\rm ann}$ is nevertheless
calculated from the model parameters. The WIMP mass is denoted by $\sigma_{\rm ann}$
and $\rho_\chi(r,z)$ is the mass
distribution function of DM particles inside the Galactic halo.
 The quantity $\xi$ parameterizes the fact that the dark halo may not be totally made of
the species under scrutiny  when this candidate possesses a relic abundance which does
not allow it to be the dominant  DM component (see e.g \cite{Bottino:2008sv} or
\cite{Bottino:1992wj}). In this case $\xi < 1$.
The quantity $g(T_{\bar p})$ in Eq.~(\ref{eq:source_pbar_prim}) denotes the
antiproton differential spectrum per annihilation event, defined as:
\begin{equation}
{\displaystyle \frac{dN_{\bar{p}}}{d E_{\bar{p}}}} \; = \; 
{\displaystyle \sum_{\rm F , h}} \, B_{\rm \chi h}^{\rm (F)} \,
{\displaystyle \frac{dN_{\bar{p}}^{\rm h}}{{d }E_{\bar{p}}}} \; .
\end{equation}
The annihilation into a quark or a gluon $h$ is realized through 
the various final states F with branching ratios $B_{\rm \chi h}^{\rm (F)}$.
Quarks or gluons may in fact be directly produced when a WIMP pair annihilates or 
they may alternatively result from the intermediate production of Higgs bosons or 
gauge bosons. Each quark or gluon $h$ then generates 
jets whose subsequent fragmentation and hadronization yield an antiproton
energy spectrum ${dN_{\bar{p}}^{\rm h}} / {d E_{\bar{p}}}$.
The single production spectra are usually evaluated within
Monte Carlo simulations of electroweak annihilation events \cite{pythia}. 
For details on the calculation of pure state ${dN_{\bar{p}}^{\rm h}} / {d E_{\bar{p}}}$
and of $g(T_{\bar p})$ in a Minimal Supersymmetric extension of the Standard 
Model (MSSM), see the Appendix of Ref. \cite{2004PhRvD..69f3501D}.

\begin{figure*}[t]
\begin{center}
{\includegraphics[width=0.55\columnwidth]{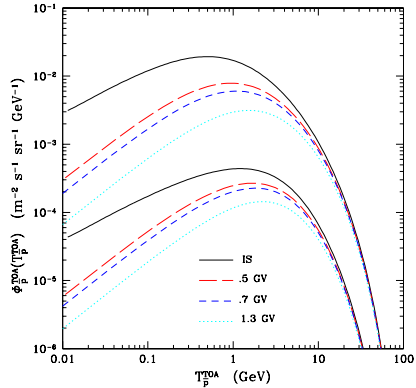}}
\end{center}
\vspace{-30pt}
\caption{Top--of--atmosphere antiproton fluxes as a function of the
antiproton kinetic energy for the $m_\chi$ = 100 GeV reference case.
The upper (lower) set of curves refer to the maximal (minimal) set of
of astrophysical parameters. Solid curves show the interstellar
fluxes. Broken curves show the effect of solar modulation at
different periods of solar activity: $\phi=500$ MV (long dashed),
$\phi=700$ MV (short dashed), $\phi=1300$ MV (dotted).
\label{fig:flux_pbar_prim_solmod}
}
\end{figure*}

The distribution of DM inside galaxies is a very debated issue.
Different analyses of rotational curves observed for several types 
of galaxies strongly favour a cored dark matter distribution, flattened towards 
the central regions (Ref. \cite{2004MNRAS.353L..17D} and references therein). 
On the other side,
many collisionless cosmological N-body simulations in $\Lambda$-CDM models
are in good agreement among themselves \cite{2007arXiv0706.1270H}, but for the
very central regions some resolution issues remain open. 
It has been recently stressed that asymptotic slopes may not be reached at
all at small scales \cite{2004MNRAS.349.1039N,2006MNRAS.365..147S,
2006AJ....132.2685M,2006AJ....132.2701G,2007ApJ...663L..53R}.
However, it is not clear whether the central cusp is steepened or flattened
when the baryonic distribution is taken into account 
(e.g. \cite{2006MNRAS.366.1529M,Mashchenko:2006dm}). A detailed discussion
may be found in Ref. \cite{Moore:thisVolume,Bullock:thisVolume,Merrit:thisVolume}. 
We wish to stress that the DM distribution is a 
crucial ingredient for the indirect signal into gamma--rays or neutrinos, while 
for antimatter it has been shown to be less relevant \cite{2004PhRvD..69f3501D,Delahaye:2007fr}, 
since antimatter diffuses and
therefore the signal at Earth is more local, less dependent on the inner structure of the
Galaxy as compared to gamma--rays and neutrinos.

To be definite, we will consider here a spherical DM galactic distribution  
with a dependence on galactocentric distance $r$ parameterized by:
\beq
\rho(r) \, = \, \rho_{\mathrm{s}} \;
\left( {\displaystyle \frac{r_{\mathrm{s}}}{r}} \right)^{\gamma} \;
\left\{
1 + \left( {\displaystyle \frac{r}{r_{\mathrm{s}}}} \right)^{\alpha}
\right\}^{{(\gamma - \beta)}/{\alpha}} \;\; .
\label{eq:DM_profile_a}
\eeq
The different profiles of Table \ref{tab_WIMP_halo_a} basically span the whole range
of reasonable halo models with respect to indirect dark matter detection prospects. 

\begin{figure*}[t]
\begin{center}
\includegraphics[width=0.49\columnwidth,bb=0 200 500 700]{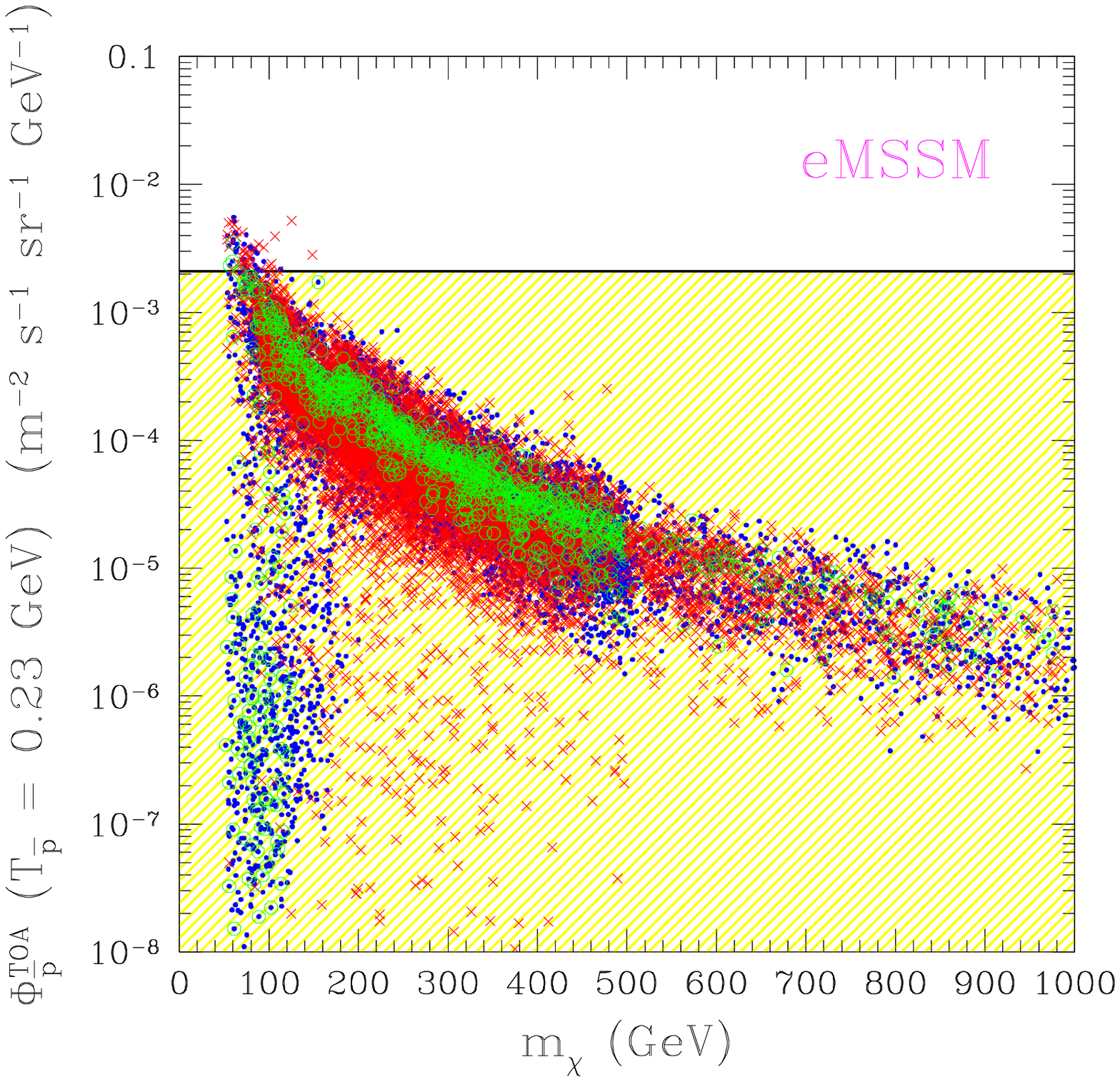}
\includegraphics[width=0.49\columnwidth,bb=0 200 500 700]{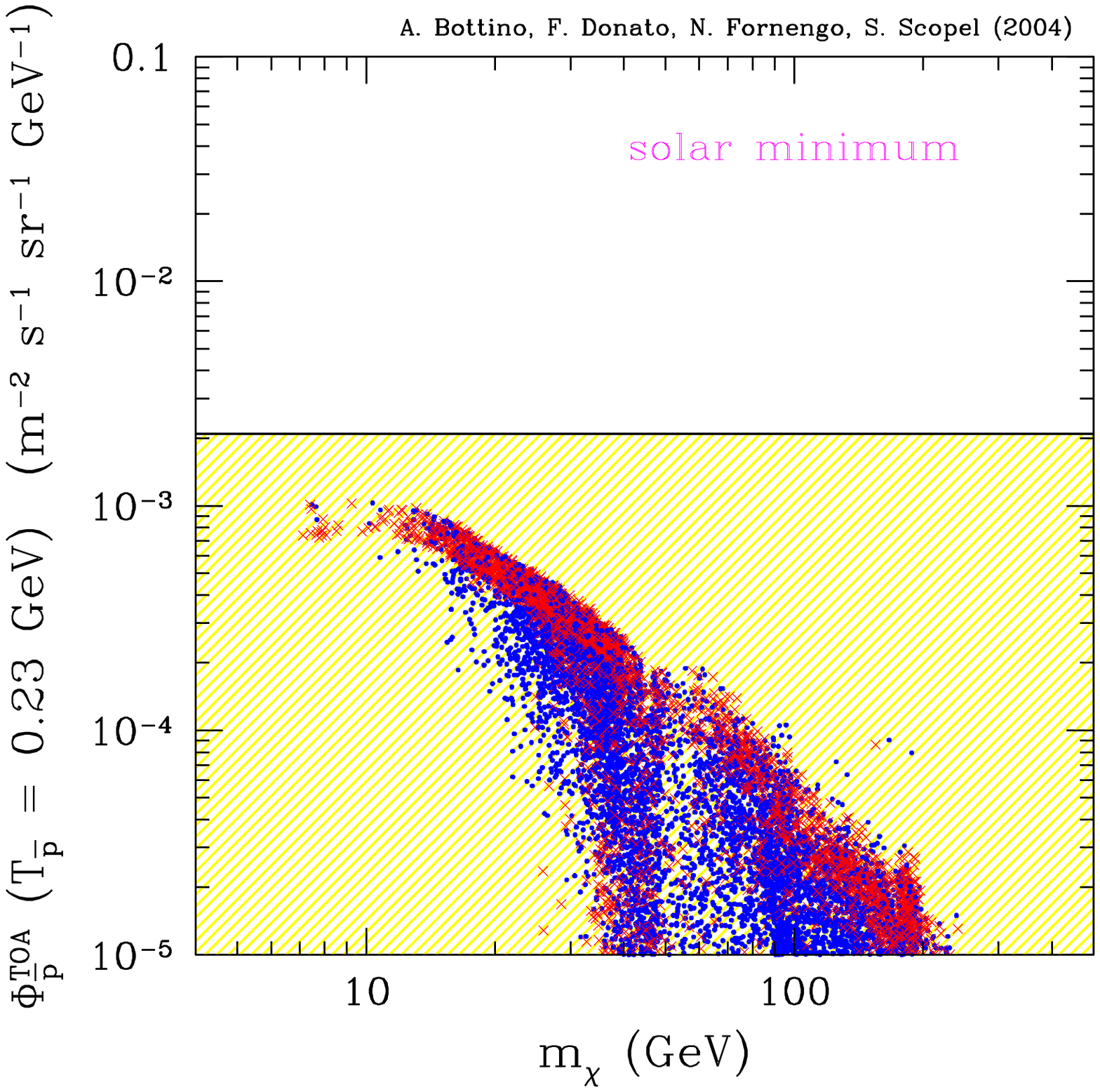}
\caption{{\sc Left:} The scatter plot shows the antiproton flux at solar minimum from
neutralino annihilation calculated at $T_{\bar p}=0.23$ GeV, as a
function of the neutralino mass for a generic scan of in a low--energy MSSM
and for the MED set of astrophysical parameters \cite{Donato:2003xg}. 
Crosses (in red) refer to cosmologically
dominant neutralinos ($0.05 \leq \Omega_{\chi} h^{2} \leq 0.3$); dots (in blue)
refer to subdominant relic neutralinos ($\Omega h^{2} < 0.05$). 
{\sc Right:} The
same as in the left panel, but calculated for in scan of supersymmetric framework
where gaugino non--universality is not assumed, and therefore lighter neutralinos
are present \cite{Bottino:2004qi}. The astrophysical parameters are set at the MIN case,
and solar modulation is at solar minimum.
}
\label{fig:pbar_neutralino}
\end{center}
\end{figure*}

The solution of the transport equation 
for a generic source term $q^{\rm prim}(r,z,E)$ within Bessel expansion 
has been derived in \cite{2002A&A...388..676B}:
\begin{eqnarray}
      N^{\bar{p},prim}_{i}(z)&=& \exp\left( \frac{V_c(|z|-L)}{2K}\right)
      \frac{y_i(L)}{A_i\sinh(S_iL/2)} \nonumber \\
      &&\left[
\cosh(S_iz/2)+\frac{(V_c+2h\Gamma^{ine}_{\bar{p}})}{KS_iA_i}\sinh(S_iz/2)\right]
      -\frac{y_i(z)}{KS_i}
\end{eqnarray}
where:
\begin{eqnarray}
         y_i(z)= 2\int_0^z\exp\left( \frac{V_c}{2K}(z-z')\right)
         \sinh\left(\frac{S_i}{2}(z-z')\right)q^{prim}_{i}(z')dz'
\end{eqnarray}
and  the quantities $S_{i}$ and $A_i$ are defined as:
\begin{equation}
S_{i} \equiv \left\{
{\displaystyle \frac{V_{c}^{2}}{K^{2}}} \, + \,
4 {\displaystyle \frac{\zeta_{i}^{2}}{R^{2}}}
\right\}^{1/2}
\;\;\; \mbox{and} \;\;\;
A_{i}(E) \equiv 2 \, h \, \Gamma^{ine}_{\bar{p}}
\; + \; V_{c} \; + \; K \, S_{i} \,
{\rm coth} \left\{ {\displaystyle \frac{S_{i} L}{2}} \right\}
\;\; .
\label{definition_Ai}
\end{equation}
In particular, at $z=0$ where fluxes are measured, we have:
\begin{equation}
     N^{\bar{p},prim}_{i}(0)= \exp\left( \frac{-V_cL}{2K} \right)
     \frac{y_i(L)}{A_i\sinh(S_iL/2)}.
     \label{eq_finale}
\end{equation}

The three propagation models featured in Table~\ref{tab_prop} have been drawn
from \cite{2004PhRvD..69f3501D}. The MED configuration provides the best fit to the B/C
measurements whereas the MIN and MAX models point respectively to the minimal and
maximal allowed antiproton fluxes which can be produced by WIMP annihilation.

Fig. \ref{fig:flux_pbar_prim_solmod} shows the TOA antiproton fluxes for 
a reference WIMP with $m_\chi=100$ GeV, $2.3 \cdot 10^{-26}$ cm$^3$ s$^{-1}$,
pure $\bar{b}b$ annihilation final state
and for the maximal and minimal sets of astrophysical parameters. 
The variation of the astrophysical 
parameters induces a much larger uncertainty on the primary than on the
secondary flux: in the first case, the uncertainty reaches
two orders of magnitude for energies $T_{\bar{p}}\lsim
1$~GeV, while in the second case it never exceeds 25\%.
The figure shows that solar
modulation has the effect of depleting the low--energy tail of the
antiproton flux. The effect is clearly more pronounced for periods of
high solar activity, when the solar wind is stronger.

Predictions for antiprotons in various realization of supersymmetric theories
have been performed (see. e.g. \cite{Donato:2003xg,Edsjo:2004pf,Bottino:2004qi,Baer:2005bu,Ferrer:2006hy,Bottino:2008mf}

. In Fig. \ref{fig:pbar_neutralino} we show the predicted antiproton
flux at kinetic energiy $T_{\bar p}=0.23$ GeV, for two different supersymmetric
models: the left panel \cite{Donato:2003xg} refers to a low--energy realization of the MSSM,
while the right panel \cite{Donato:2003xg} stands for a scan of a supersymmetric scheme where gaugino
non--universality is not assumed and therefore light neutralinos are present \cite{Bottino:2002ry,Bottino:2003iu}.
In both cases the shaded (yellow) region denotes the amount of antiprotons, in excess
of the secondary component , which can be accommodated at
$T_{\bar p}=0.23$ GeV in order not to exceed the observed flux, as measured
by BESS. All the points of the scatter plot that lie
below the horizontal black line are therefore compatible with observations.

\section{Antideuterons in Cosmic Rays}
\label{sec:dbars}
In the seminal Paper \cite{2000PhRvD..62d3003D}, it was proposed   to look for
cosmic antideuterons  (\dbar) as a possible indirect signature for galactic  dark matter.
It was shown that the  antideuteron spectra deriving from DM annihilation  is
expected to be much flatter than the  standard astrophysical component at low
kinetic energies, $T_{\overline{d}}\lsim$ 2-3 GeV/n. 
This argument motivated the proposal of a new space-borne experiment
\cite{2002ApJ...566..604M,2004NIMPB.214..122H,2006JCAP...01..007H} looking for
cosmic antimatter (antiproton and antideuteron) and having the potential to
discriminate between standard and exotic components for a wide range of DM
models. 
The present experimental upper limit \cite{2005PhRvL..95h1101F} is still far
from the expectations on the secondary antideuteron flux which are produced by 
spallation of cosmic rays on the interstellar medium 
\cite{1997PhLB..409..313C,Duperray:2005si}, but  perspectives for the near
future are very encouraging.

Once the astrophysical framework for the transport of (anti)nuclei is set,  the calculation
of the antideuteron flux rests on the \dbar\ specificities regarding the source term 
(i.e. primary or secondary) and its
nuclear interactions [r.h.s. of Eq.~(\ref{master_equation})].
The tertiary term (see Eq. \ref{tertiary}) requires specific inelastic non-annihilating
cross sections which are  detailed in the Appendix of Ref. \cite{2008PhRvD..78d3506D}.

The production of cosmic antideuterons is based on the fusion process of a \pbar\
and \nbar\ pair. One of the simplest but powerful treatment of the fusion of two or
more nucleons is based on the so--called coalescence model which, despite its
simplicity, is able to reproduce remarkably well the {available} data on light nuclei and
antinuclei production in different kinds of collisions. 
In the coalescence model, the momentum distribution of the (anti)deuteron
is proportional to the product of the (anti)proton and (anti)neutron momentum
distribution \cite{1963PhRv..129..836B,1963PhRv..129..854S}.
That function depends on the difference $\Delta_{\vec{k}}$
between (anti)nucleon momenta. It is strongly peaked around
$\Delta_{\vec{k}}\simeq \vec{0}$ (compare the minimum energy to form
a \dbar, i.e. $4m_p$, with the binding energy $\sim 2.2$~MeV), so that
\begin{equation}
\vec{k}_{\bar{p}}\simeq \vec{k}_{\bar{n}} \simeq \frac{\vec{k}_{\bar{D}}}{2}\;.
\end{equation}
The \dbar\ density in momentum space is thus written as
the \pbar\ density times the probability to find an $\bar{n}$
within a sphere of radius $p_0$ around $\vec{k}_{\bar{p}}$
(see, e.g. Ref.~\cite{1994ZPhysC..61...683}):
\begin{equation}
\gamma \frac{d{\cal N}_{\bar{D}}}{d\vec{k}_{\bar{D}}} = \frac{4\pi}{3} p_0^3
  \cdot    \gamma \frac{d{\cal N}_{\rm \bar{p}}}{d\vec{k}_{\bar{p}}}
  \cdot    \gamma \frac{d{\cal N}_{\rm \bar{n}}}{d\vec{k}_{\bar{n}}}\;.
  \label{eq:coal}
\end{equation}
The coalescence momentum $p_0$ is a free parameter
constrained by data on hadronic production
In \cite{Duperray:2005si}, a large set of data is used,
including many $pA$ reactions (see their Tab.~I and
references therein). This leads to an estimate of $p_0=79$~MeV.
At LEP energies, (anti)deuteron production occurs through $e^+e^-$ annihilations into
$q\bar{q}$ pairs, an electroweak
 mechanism similar to the \dbar\ production in DM annihilation reactions.
Based on theoretical arguments, it has been argued \cite{1994ZPhysC..61...683}
that the antideuteron  yields in $e^+e^-$ reactions should be smaller than in hadronic reactions.
However, the ALEPH Collaboration \cite{2006PhLB..639..192A} has found that this theoretical
prediction (see Fig.~5 in ALEPH paper) underestimates their measured \dbar\ inclusive cross
section. They derive (see their Fig.~6)  a value $B_2=3.3\pm 0.4\pm0.1 \times 10^{-3}$~GeV$^2$
at the $Z$ resonance, which translates into $p_0=71.8\pm 3.6$~MeV, very close to
the $p_0=79$~MeV derived for the hadronic production.
From a rough estimate of the fusion process, one can estimate that 
the antideuteron fluxes are a factor of $10^4$ lower than the
antiproton ones. We will see in the following, and by comparing the results of the 
previous Section, that this result is recovered for both the 
primary and the secondary \dbar\ fluxes at Earth.

\subsection{Secondary Antideuterons}
\label{dbar_sec}

\begin{figure*}[t]
\begin{center}
\vspace{-1cm}
\includegraphics[width=0.53\columnwidth]{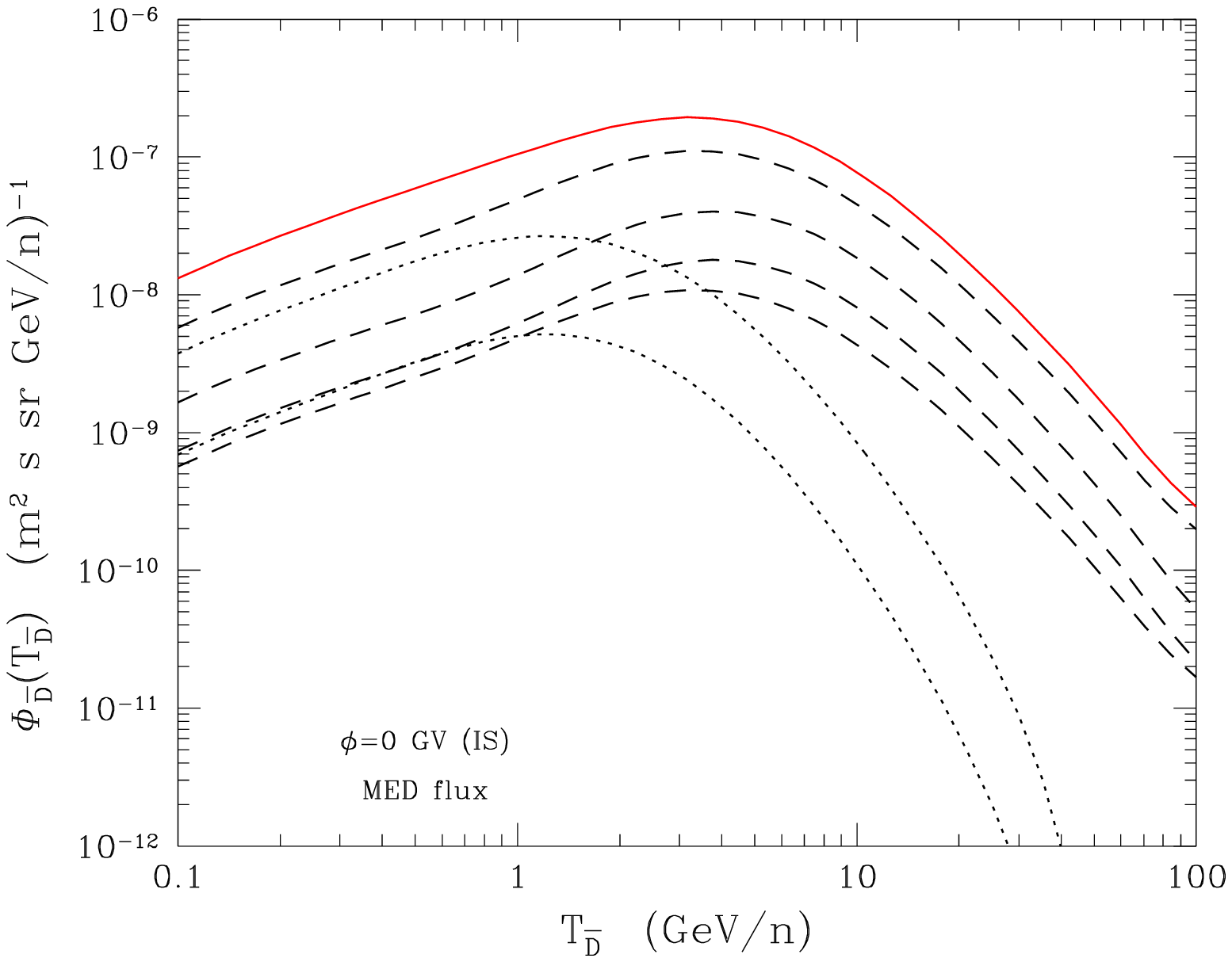}
\hspace{-1cm}
\includegraphics[width=0.53\columnwidth]{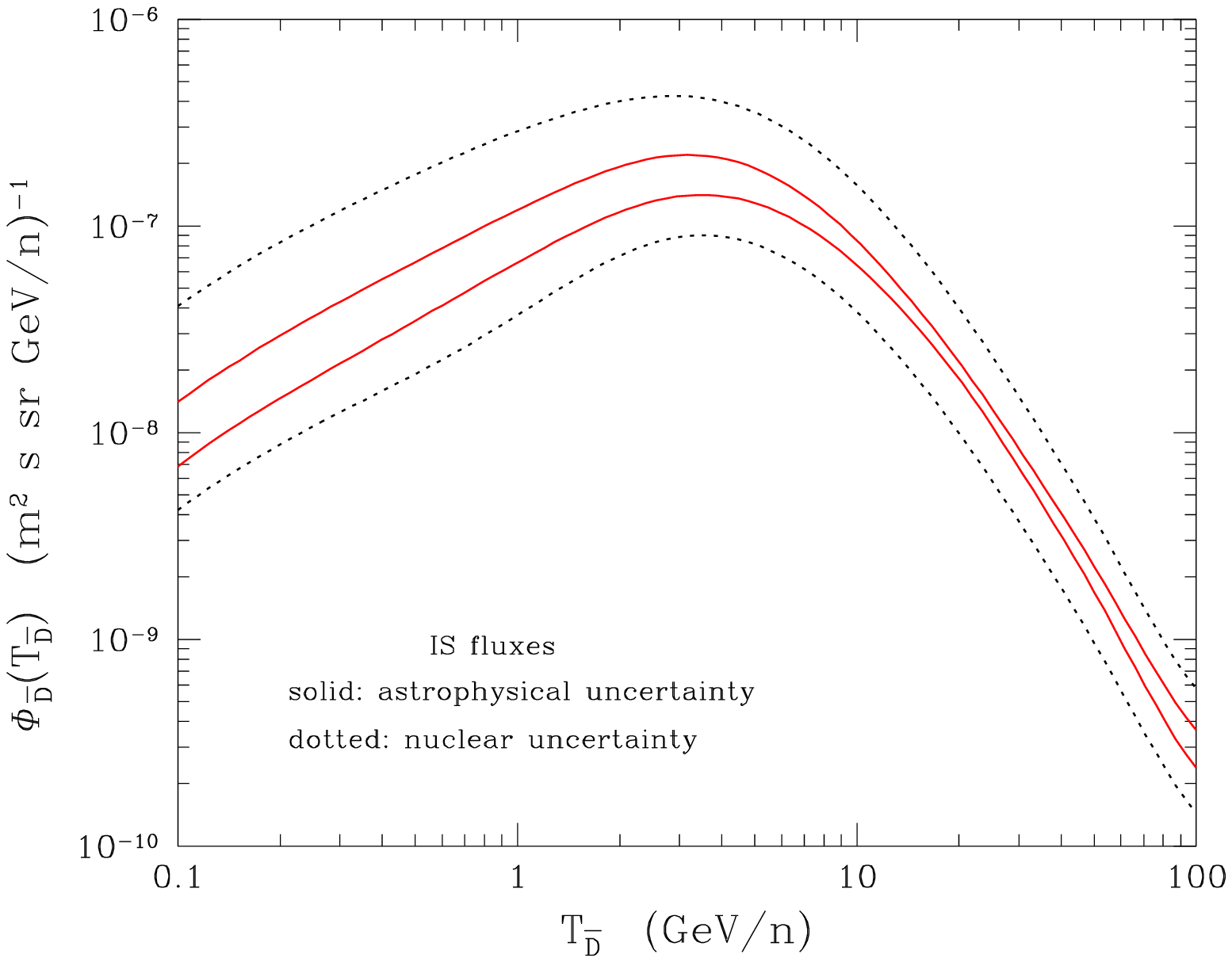}
\end{center}
\vspace{-1cm}
\caption{Left panel: 
Contribution of all nuclear channels to the \dbar\  secondary flux. Dashed
lines, from top to bottom refer to: p+H, p+He, He+H, He+He. Dotted lines, from top to
bottom stand for: \pbar+H, \pbar+He. Solid line: sum of all the components.
Right panel: Dominant uncertainties on the interstellar secondary \dbar\ flux. 
Solid lines: propagation uncertainty band. 
Dotted lines: nuclear uncertainty band. }
\label{fig:dbar_sec}
\end{figure*}

The secondary \dbar\ flux is the sum of the six contributions corresponding to $p$, He
and \pbar\ cosmic ray fluxes impinging on H and He IS gas (other reactions are
negligible \cite{Duperray:2005si}).
Contributions to the \dbar\ flux from $\bar{p}+H$ and $\bar{p}+{\rm He}$
reactions are evaluated using the \pbar\ flux calculated in the same
run. The production cross sections for these specific processes
are those given in Ref. \cite{Duperray:2005si}.
The solution to the propagation equation has the same expression as for secondary
antiprotons \cite{Donato:2008yx}.

The different contributions to the total secondary antideuteron flux, 
calculated for the best fit propagation configuration 
(the ``MED" one in Table \ref{tab_prop}), 
are shown in the left panel of Fig.~\ref{fig:dbar_sec}. 
As expected, the dominant production channel is the one from $p$-H collisions,
followed by the one from cosmic protons on IS helium (p-He).  
As shown in Ref. \cite{Duperray:2005si}, the \pbar+H channel is dominant at low energies,
and negligible beyond a  few GeV/n. The effect of
energy losses, reacceleration and  tertiaries add up to replenish the low
energy tail. The maximum of the total 
flux reaches the value of  $2\cdot 10^{-7}$  particles 
(m$^2$ s sr GeV/n)$^{-1}$ at 3-4 GeV/n. At 100 MeV/n it is decreased by 
an order of magnitude, thus preserving an  interesting window for possible
exotic contributions characterized by a flatter spectrum (see next Section).

For the determination of the propagation uncertainties, one can 
calculate the secondary antideuteron flux for all the propagation parameter 
combinations providing an acceptable fit to stable nuclei 
\cite{2001ApJ...555..585M}. 
 The resulting envelope for the secondary antideuteron flux 
 is presented in Fig.~\ref{fig:dbar_sec}-left.
 The solid lines  delimit the uncertainty band due to the
degeneracy of the propagation parameters: at energies below 1--2 GeV/n, the uncertainty 
is 40-50 \% around the average flux, while at 10 GeV/n it decreases to 
 $\sim 15$ \%. This behavior is analogous to that obtained for \pbar\ \cite{2001ApJ...563..172D}
and is easily understood. The degenerate transport parameters combine to give
the same grammage in order to reproduce the B/C ratio. Indeed, the 
grammage crossed by C to produce the secondary species B is also
crossed by $p$ and He to produce the secondary \pbar\ and \dbar. In short,
a similar propagation history associated with a well constrained B/C
ratio explains the small uncertainty. 
The possible nuclear uncertainty can arise from two different
sources. The first one is directly related to the elementary production process
$d\sigma^{\rm R}_{\rm \bar{p}}$ and has been estimated to be conservatively 
$\pm 50 \%$ \cite{2008PhRvD..78d3506D}. Second, there is the
uncertainty on the coalescence momentum $p_0$.  Using an independent model (i.e.
different from the coalescence scheme) for
\dbar\ production, Ref.~\cite{Duperray:2005si} found that, conservatively, the
\dbar\ background was certainly no more than twice the flux calculated with
$p_0=79$~MeV.
The dotted lines in Fig.~\ref{fig:dbar_sec}, right panel,  take into account the
sum of all the possible uncertainties of nuclear source
\cite{2008PhRvD..78d3506D}.  At the lowest energies the flux is uncertain by
almost one order of magnitude, at 100 GeV/n by a factor of 4.

\subsection{Antideuterons from DM annihilation}
\label{dbar_prim}
The source term for primary \dbar\ has the same form as the one 
for primary antiprotons in Eq. \ref{eq:source_pbar_prim} 
($\bar{p}\rightarrow\bar{D}$). 
The production of antideuterons from the pair-annihilation of 
supersymmetric dark matter
particles in the halo of our Galaxy was proposed in \cite{2000PhRvD..62d3003D}, and
subsequently discussed in \cite{2005JCAP...12..008B} also for universal
extra-dimension, Kaluza-Klein and warped extra-dimension dark matter models. 
In ref. \cite{2008PhRvD..78d3506D} primary antideueterons in different supersymmetric
scenarios have been calculated in the full 2D propagation model, with a thorough
estimation of all the possible uncertanty sources.  
As previously discussed (see
Sect.~\ref{sec:dbars}) the production  of a \dbar\ relies on the availability of a
\pbar\ -- \nbar\ pair in a single  DM annihilation.  As in
Ref.~\cite{2000PhRvD..62d3003D},
we assume that the probability to form an antiproton (or an antineutron)
with momentum $\vec{k}_{\bar{p}}$ ($\vec{k}_{\bar{n}}$), is essentially isotropic:
\beq
{\displaystyle \frac{dN_{\bar{p}}}{d E_{\bar{p}}}}(\chi + \chi \to \bar{p} + \ldots)
\; = \;
4 \pi \, k_{\bar{p}} \, E_{\bar{p}} \, {\cal F}_{\bar{p}}(\sqrt{s} = 2 m , E_{\bar{p}})
\; .
\eeq
Applying the factorization--coalescence scheme discussed above leads
to the antideuteron differential multiplicity
\beq
{\displaystyle \frac{dN_{\bar{D}}}{d E_{\bar{D}}}} = 
\left( {\displaystyle \frac{4 \, p_0^{3}}{3 \, k_{\bar{D}}}} \right)
\cdot
\left( {\displaystyle \frac{m_{\bar{D}}}{m^2_{\bar{p}}}} \right)
\cdot
{\displaystyle \sum_{\rm F , h}}  B_{\rm \chi h}^{\rm (F)} 
\left\{
{\displaystyle \frac{dN_{\bar{p}}^{\rm h}}{d E_{\bar{p}}}}
\left( E_{\bar{p}} = \frac{E_{\bar{D}}}{2} \right)
\right\}^{2} \; .
\label{dNdbar_on_dEdbar_susy}
\eeq
We assume, as discussed in Sect.~\ref{sec:dbars}, that the same value 
of the coalescence momentum $p_0=79$ MeV holds as for hadronic reactions.

\begin{figure*}[t]
\begin{center}
\includegraphics[width=0.55\columnwidth]{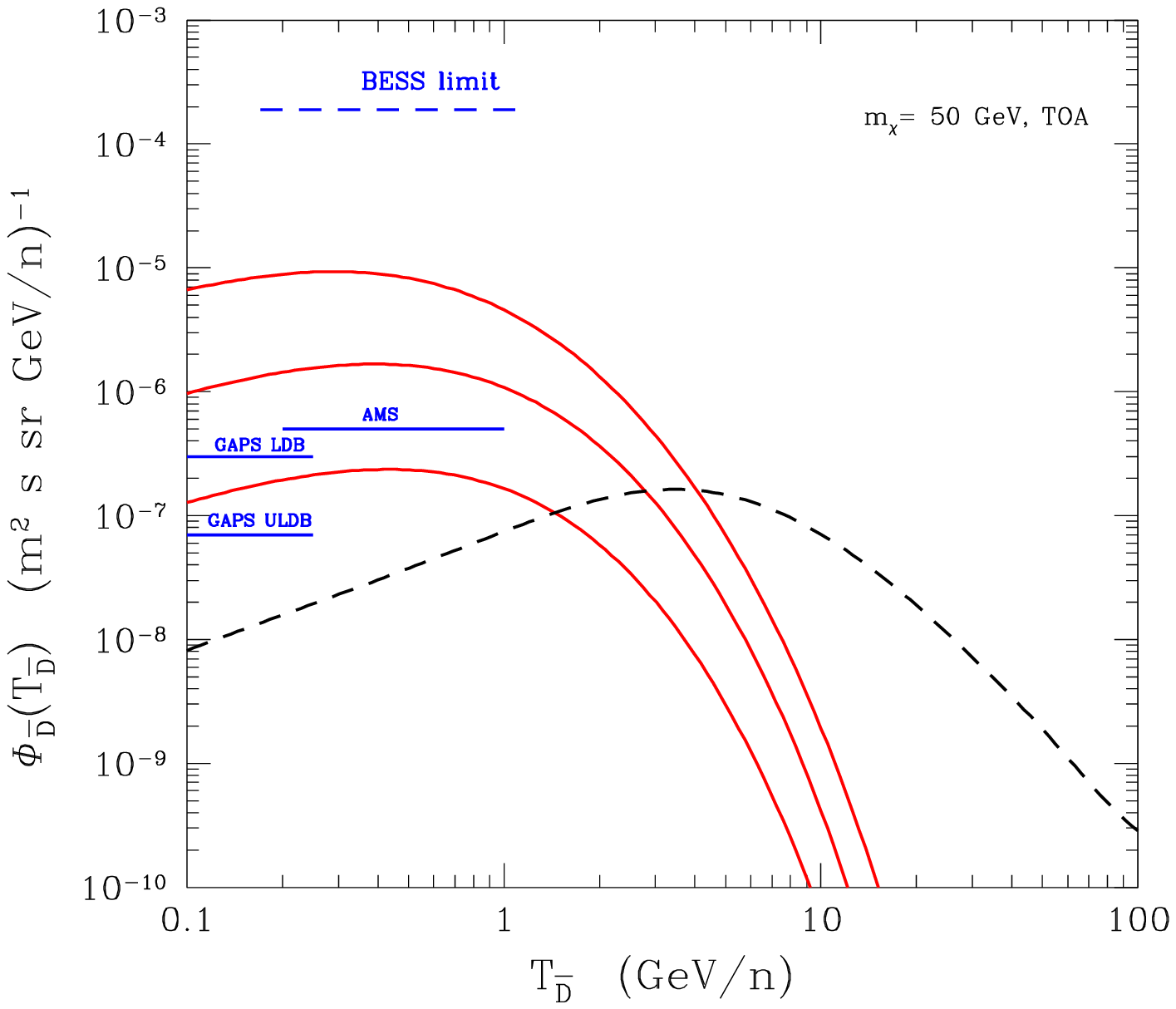}
\hspace{-1.5cm}
\includegraphics[width=0.55\columnwidth]{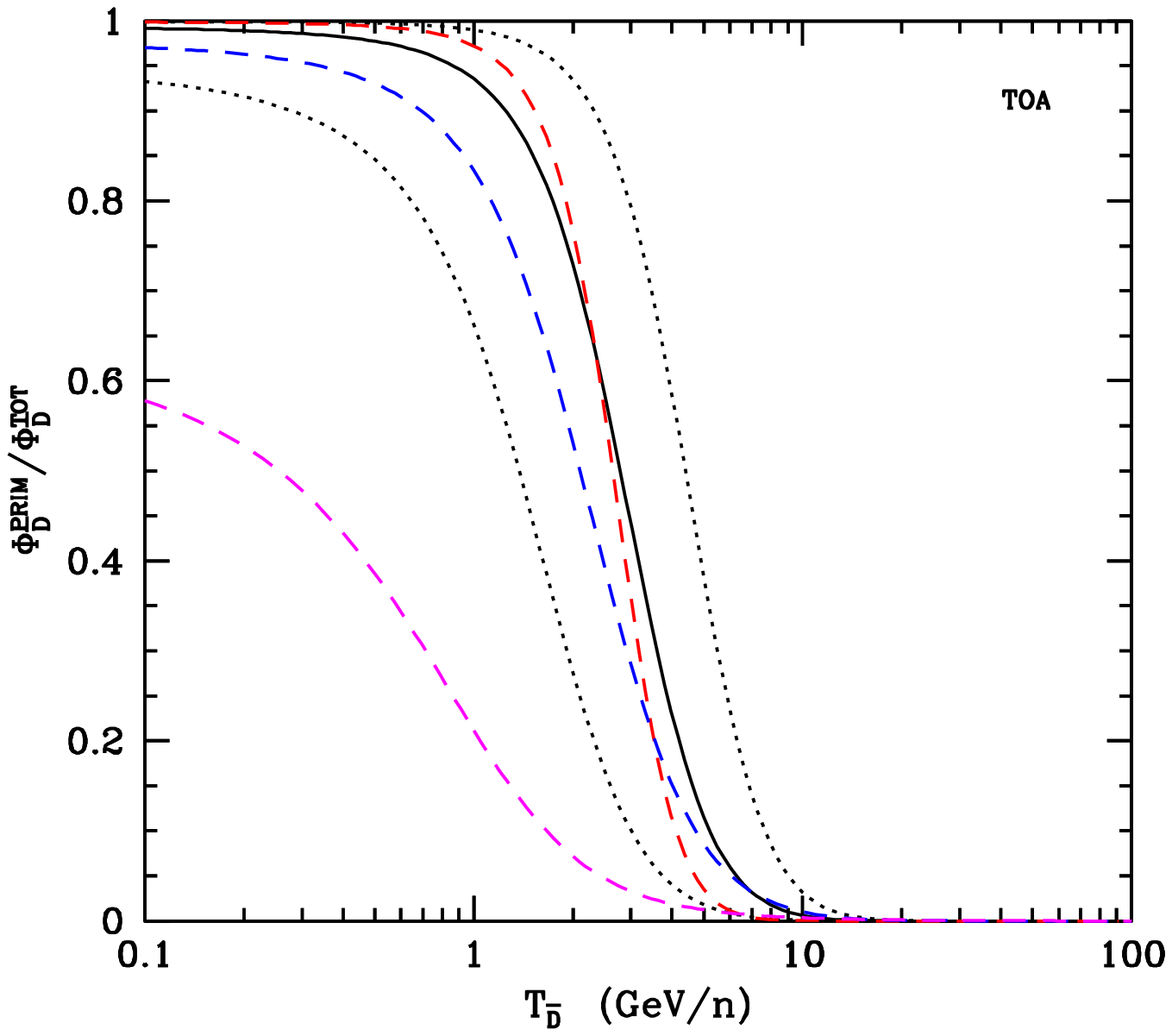}
\end{center}
\vspace{-1cm}
\caption{Left panel: TOA primary (red solid lines) and secondary (black dashed line) 
antideuteron fluxes, modulated at solar minimum. The signal is derived for a 
$m_\chi$=50 GeV WIMP and for the three propagation models of Table \ref{tab_prop}. 
The secondary flux
is shown for the median propagation model. The upper dashed horizontal line shows the
current BESS upper limit on the search for cosmic antideuterons. The 
three horizontal solid (blue) lines are the estimated sensitivities for (from
top to bottom): AMS--02, GAPS on a long (LDB) and
ultra--long (ULDB) duration balloon flights \cite{2004NIMPB.214..122H,2006JCAP...01..007H,koglin_taup}.
Right panel: Ratio of the primary to total (signal+background) TOA antideuteron flux. 
Solid (black) curve refers to a WIMP mass of $m_\chi$=50 GeV and for the MED 
propagation parameters. Dotted (black) lines show the MAX (upper) and MIN (lower) cases. 
Dashed lines refer to the MED propagation parameters and different masses, which
are (from top to bottom): $m_\chi$=10, 100, 500 GeV (red, blue, magenta respectively).}
\label{fig:dbar_prim}
\end{figure*}

The solution to the diffusion equation is provided by
Eq. \ref{eq_finale} where the specific cross sections have been 
reviewd, for instance, in Ref. \cite{2008PhRvD..78d3506D}. 
In the left panel of Fig. \ref{fig:dbar_prim} the 
secondary \dbar\ flux for the median
configuration of Table \ref{tab_prop} is plotted  alongside the
primary flux from $m_\chi=$50 GeV, calculated for the maximal, median
and minimal  propagation scenarios.  The present BESS upper limit on
the (negative) antideuteron search \cite{2005PhRvL..95h1101F}  is at a
level of 2$\cdot 10^{-4}$ (m$^2$ s sr GeV/n)$^{-1}$. We also plot the
estimated sensitivities of the gaseous antiparticle spectrometer GAPS
on a long duration balloon flight (LDB) and  an ultra--long duration balloon
mission (ULDB) 
\cite{2004NIMPB.214..122H,2006JCAP...01..007H,koglin_taup}, 
and of AMS--02 for three years of data taking. 
 The perspectives to explore a part of  the region
where DM annihilation are mostly expected 
(i.e. the low--energy tail) are  very promising.  If one of
these experiments will measure at least 1 antideuteron, it will be a
clear  signal of an exotic contribution to the cosmic antideuterons.

The discrimination power between primary and secondary \dbar\ flux may be deduced
from  the right panel of Fig.~\ref{fig:dbar_prim}.  The ratio of the primary to
total TOA \dbar\ flux is plotted as a function  of the kinetic energy per nucleon,
for the three representative propagation models and different WIMP masses {(the
annihilation cross section is again fixed at the reference value)}. This ratio keeps
higher than 0.7 for $T_{\bar{D}}<1$ GeV/n except for $m_\chi=$500 GeV.  For
propagation models with $L \gsim 4$ kpc -- which is a very reasonable expectation --
this ratio   is at least 0.9 for masses below 100 GeV. Increasing the WIMP mass, we
must descend to lower  energies in order to maximize the primary--to--secondary
ratio.  However, for a  $m_\chi=$500 GeV WIMP we still have a 50-60\% of DM
contribution in the 0.1-0.5 GeV/n range. Of course, the evaluation of the
theoretical uncertainties presented in this Paper must be  kept in mind while
confronting to real data.  Fig.~\ref{fig:dbar_prim} clearly states that the antideuteron
indirect DM detection technique is probably  the most powerful one for low and
intermediate WIMP--mass haloes.

The uncertainties due to propagation are similar to the ones quoted for primary antiprotons
\cite{2008PhRvD..78d3506D}. At the lowest energies of  hundreds of MeV/n the total uncertainty reaches
almost 2 orders of magnitude,  while at energies above 1 GeV/n it is about a factor of 30. 

\begin{figure*}[t]
\begin{center}
\includegraphics[width=0.49\columnwidth]{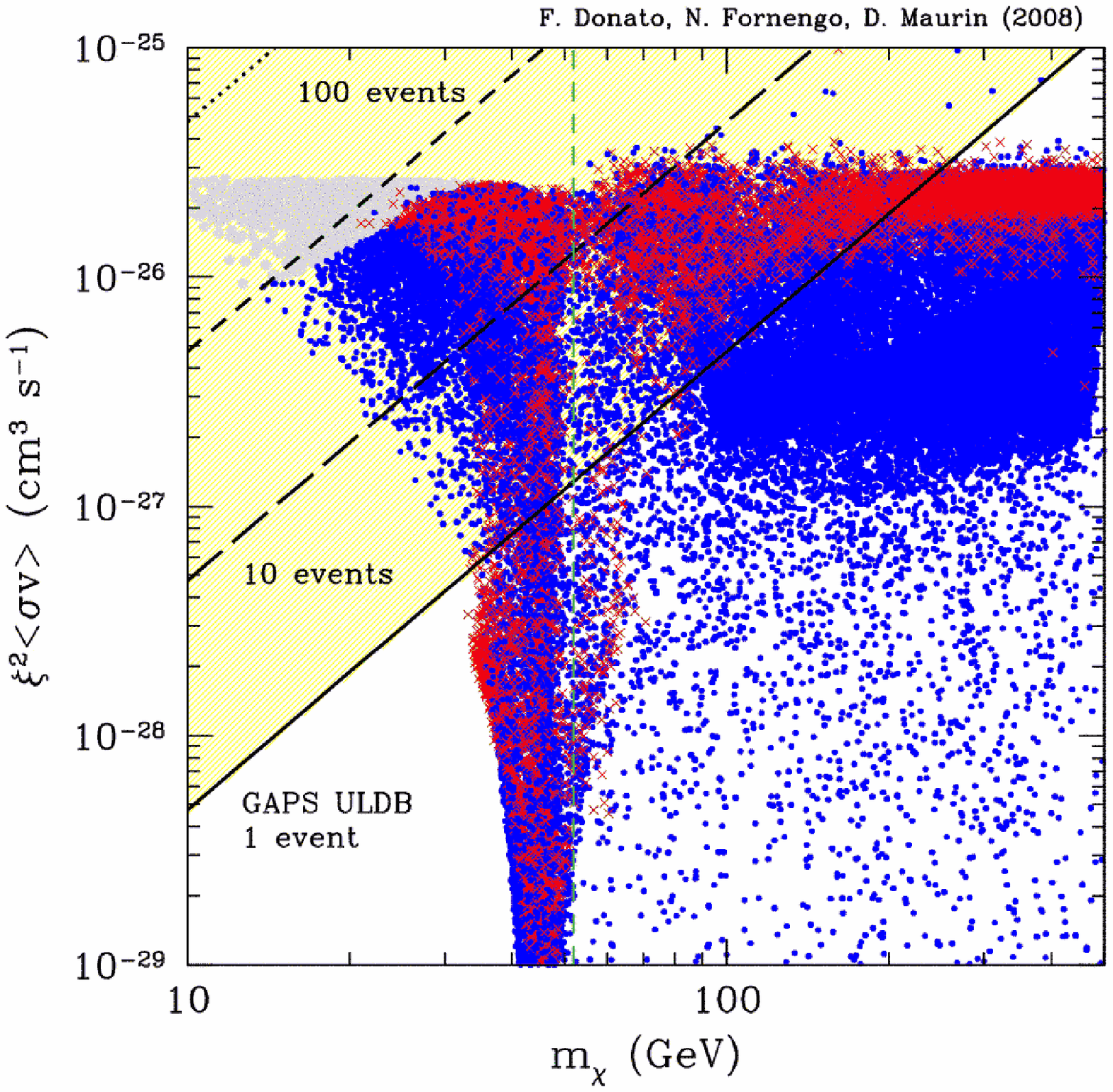}
\includegraphics[width=0.49\columnwidth]{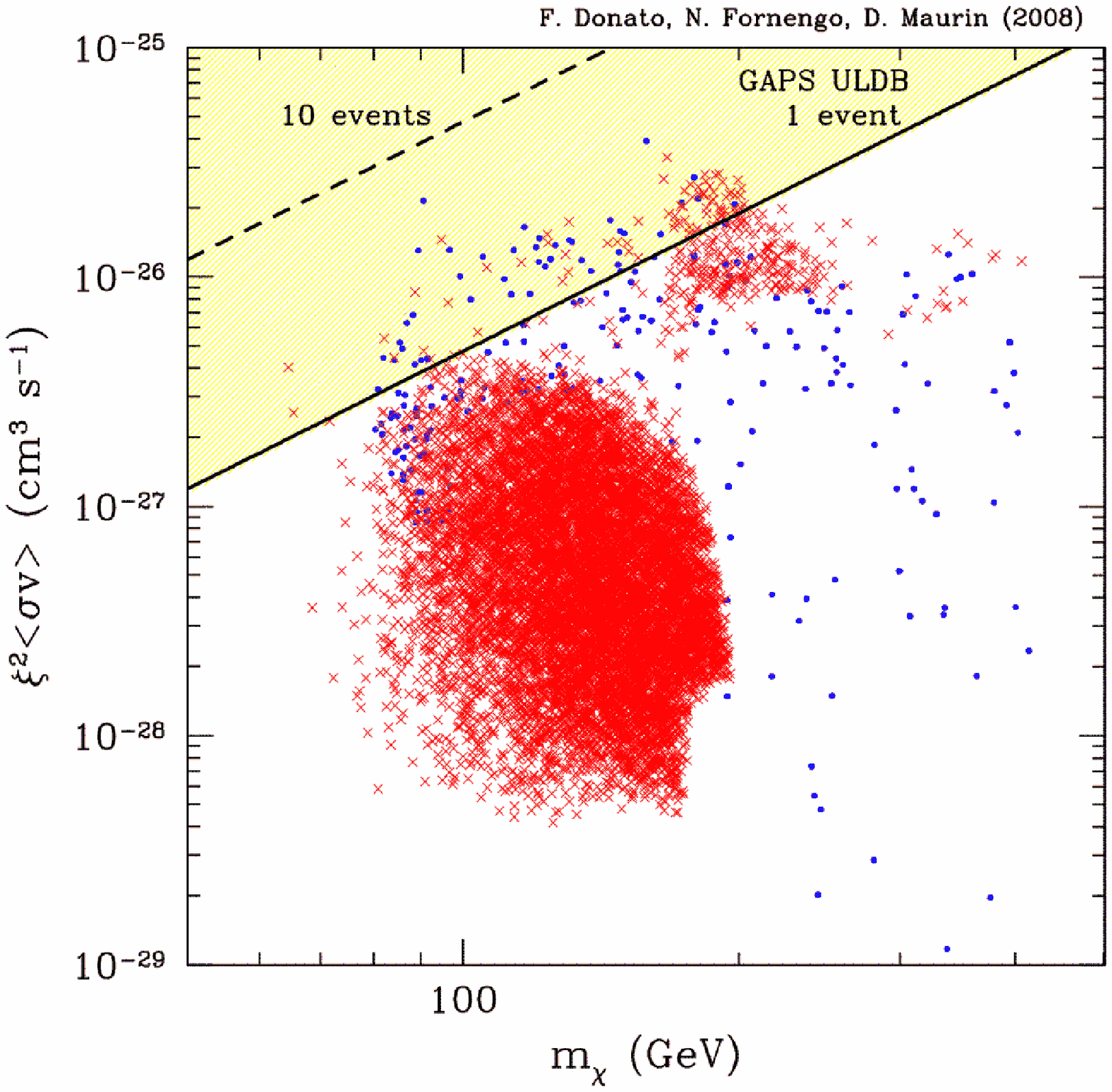}
\caption{{\sc Left:} GAPS Ultra Long Duration Flight (ULDF) reach compared to
predictions for neutralino dark matter in low--energy supersymmetric models,
shown in the plane effective annihilation cross section $\xisigmav_0$ vs.
neutralino mass
$m_\chi$ \cite{2008PhRvD..78d3506D}. The solid, long--dashed and short--dashed
lines show our estimate for the capability of GAPS ULDB
of measuring 1, 10 and 100 events, respectively, for the MED propagation model.
The scatter plot reports the quantity $\xisigmav_0$ calculated
in a low--energy MSSM (for masses above the vertical [green] dashed line) and
in non--universal gaugino
models which predict low--mass neutralinos \cite{Bottino:2003iu,Bottino:2002ry}.
[Red] Crosses refer to cosmologically
dominant neutralinos, while [blue] dots stand for subdominant neutralinos.
Grey point are excluded by antiproton searches.
{\sc Right:} The same as in the left panel, but for SUGRA scheme with
non--universality in the higgs sector \cite{2008PhRvD..78d3506D}.
}
\label{fig:dbar_neutralino}
\end{center}
\end{figure*}

Examples for theoretical calculations for neutralino dark matter are shown in Fig. 
\ref{fig:dbar_neutralino}, where the reaching capabilities of the GAPS detector are derived and
shown in terms of the neutralino annihilation parameters. The shaded band denotes the region
which will be covered by GAPS with a Ultra Long Duration Flight: sensitivity to neutralinos
from a mass of a few GeV up to few hundreds of GeV is at hand and it will cover a large
portion of the supersymmetric parameter space for models with gaugino non--universality
\cite{Bottino:2003iu,Bottino:2002ry} or low--energy MSSM \cite{2008PhRvD..78d3506D}.

%
%
\section{Positrons in Cosmic Rays}
In the case of positrons and electrons, the master equation~(\ref{master_equation})
describing the propagation of cosmic rays throughout the DH is dominated by
space diffusion and energy losses. Above a few GeV, synchrotron radiation in the
galactic magnetic fields as well as inverse Compton scattering on stellar light
and on CMB photons dominate, hence the positron loss rate
\beq
b^{\rm loss}(E) \, = \, \left\langle \dot{E} \right\rangle \, = \, - \,
{\displaystyle \frac{E^{2}}{E_{0} \tau_{E}}} \;\; .
\label{positron_loss_rate}
\eeq
The energy of reference $E_{0}$ is set equal to 1 GeV while the typical energy loss
time $\tau_{E}$ is of order $10^{16}$ s. The master equation for positron propagation
simplifies into
\beq
- \, K  \, \Delta \psi \, + \, \partial_{E} \! \left\{ b^{\rm loss}(E) \, \psi \right\}
\, = \, q \left( {\mathbf x} , E \right) \;\; .
\label{master_positron_1}
\eeq
Above a few MeV, positrons are ultra--relativistic and the rigidity ${\mathcal R}$
is proportional to the energy $E$. The space diffusion coefficient boils down
to $K(\epsilon) = K_{0} \, {\epsilon}^{\delta}$ where $\epsilon = {E}/{E_{0}}$.
%
%
%
The solution of equation~(\ref{master_positron_1}) proposed
by~\cite{1974Ap&SS..29..305B,Baltz:1998xv} is based on replacing
the energy $E$ by the pseudo--time
\begin{equation}
\tilde{t}(E) \, = \, \tau_{E} \;
\left\{
v(E) \, = \, {\displaystyle \frac{\epsilon^{\delta - 1}}{1 - \delta}}
\right\} \;\; ,
\label{connection_E_pseudo_t}
\end{equation}
and leads to the well--known heat equation
\begin{equation}
{\displaystyle \frac{\partial \tilde{\psi}}{\partial \tilde{t}}} \; - \;
K_{0} \; \Delta \tilde{\psi} \, = \,
\tilde{q} \left( {\mathbf x} , \tilde{t} \, \right) \;\; ,
\label{master_positron_3}
\end{equation}
where the space and energy positron density is now given by
$\tilde{\psi} = \epsilon^{2} \, \psi$ whereas the positron production rate
has become
$\tilde{q} = \epsilon^{2 - \delta} \, q$. Because $\epsilon$ is dimensionless,
both $\tilde{\psi}$ and $\tilde{q}$ have the same dimensions as before.
In this formalism, the energy losses which positrons experience are described
by an evolution of their density with respect to the pseudo--time $\tilde{t}$.
Equipped with these notations, we can write the cosmic ray positron density
as the convolution
{\small
\begin{equation}
\psi_{e^{+}} \!\! \left( {\mathbf x} , E \right) \, = \,
{\displaystyle \int_{E_{S} = E}^{E_{S} = + \infty}} dE_{S} \;
{\displaystyle \int}_{\rm \!\! DH} \!\! d^{3}{\mathbf x}_{S} \;\;
G_{e^{+}} \!\! \left( {\mathbf x} , E \leftarrow {\mathbf x}_{S} , E_{S} \right) \;
q_{e^{+}} \!\! \left( {\mathbf x}_{S} , E_{S} \, \right) \;\; .
\label{psi_convolution_positron}
\end{equation}}
The positron propagator
$G_{e^{+}} \!\! \left( {\mathbf x} , E \leftarrow {\mathbf x}_{S} , E_{S} \right)$
is defined as the probability for a particle injected at $\mathbf{x}_{S}$ with the
energy $E_{S}$ to reach the location $\mathbf{x}$ with the degraded energy
$E \leq E_{S}$.
It is proportional to the Green function $\tilde{G}$ of the
heat equation~(\ref{master_positron_3}) through
\begin{equation}
G_{e^{+}} \!\! \left( {\mathbf x} , E \leftarrow {\mathbf x}_{S} , E_{S} \right) \, = \,
{\displaystyle \frac{\tau_{E}}{E_{0} \, \epsilon^{2}}} \;
\tilde{G} \left( {\mathbf x} , \tilde{t} \leftarrow {\mathbf x}_{S} , \tilde{t}_{S} \right)
\;\; ,
\label{positron_propagator}
\end{equation}
where the connection between the energy $E$ and pseudo--time $\tilde{t}$ is given by
relation~(\ref{connection_E_pseudo_t}).

%
%
The build up our intuition, it is useful to derive the heat Green function
$\tilde{G}$ connecting a source at $\mathbf{x}_{S}$ to the Earth in the
simple case of an infinite DH. In this limit, there are no vertical boundaries
and we get the Gaussian distribution
\begin{equation}
\tilde{G} \left( {\mathbf x}_{\odot} , \tilde{t} \leftarrow \mathbf{x}_{S} , \tilde{t}_{S} \right)
\, = \,
\left\{ \frac{1}{4 \, \pi \, K_{0} \, \tilde{\tau}} \right\}^{3/2} \,
\exp \left\{ - \,
{\displaystyle \frac{r_{\oplus}^{2}}{4 \, K_{0} \, \tilde{\tau}}} \right\} \;\; ,
\label{propagator_reduced_3D_lD_a}
\end{equation}
where $\tilde{\tau} = \tilde{t} - \tilde{t}_{S}$ is the typical duration
over which the positron energy decreases from $E_{S}$ to $E$. That timescale
also includes information on the diffusion process. The distance between the
source at $\mathbf{x}_{S}$ and the Earth is denoted by $r_{\oplus}$.
The concept of positron horizon is based on the Gaussian
distribution~(\ref{propagator_reduced_3D_lD_a}) which is roughly constant within
a sphere of radius
\beq
\lD \, = \, \sqrt{4 K_{0} \tilde{\tau}} \;\; ,
\label{definition_lD}
\eeq
and decreases sharply outside. So does the positron Green function $G_{e^{+}}$.
The positron sphere -- whose center is at the Earth where the observer stands
-- delineates actually the region of the diffusive halo from which positrons
predominantly originate. The typical diffusion length $\lD$ gauges how far
particles produced at the energy $E_{S}$ travel before being detected with
the energy $E$. It encodes at the same time the energy loss process and the
diffusion throughout the magnetic fields of the Galaxy. A rapid inspection of
equation~(\ref{connection_E_pseudo_t}) shows that $\lD$ increases as the detected
energy $E$ decreases, except for energies $E_{S}$ at the source very close to $E$.
The positron sphere is indeed fairly small at high energies, say above $\sim$ 100 GeV,
whereas it spreads over several kiloparsecs below 10 GeV. In the case where
$E_{S} = 100$ GeV for instance, $\lD$ exceeds 3~kpc below an energy $E$ of
$\sim 8$ GeV.

%
%
The diffusive halo inside which cosmic rays propagate before escaping into the
intergalactic medium is actually finite and $\tilde{G}$ should account for
that effect.
In spite of the boundaries at $r = R \equiv 20$ kpc, we can decide that
cosmic ray diffusion is not limited along the radial direction and that
it operates as if it took place inside an infinite horizontal slab with
half--thickness $L$. Sources located beyond the galactic radius $R$ are
disregarded since the convolution~(\ref{psi_convolution_positron})
is performed only over the DH. Because their energy is rapidly degraded as
they propagate, positrons are produced close to where they are observed.
Neglecting the effect of radial boundaries on the propagator $G_{e^{+}}$
turns out to be a fair approximation \cite{2008PhRvD..77f3527D} because
positrons do not originate from far away on average. The effects of the
radial boundaries down at the Earth are not significant insofar as cosmic
rays tend to leak above and beneath the diffusive halo at $z = \pm L$ instead
of traveling a long distance along the galactic plane.

The infinite slab hypothesis allows the radial and vertical directions
to be disentangled and to express the reduced propagator $\tilde{G}$ as
\begin{equation}
\tilde{G}
\left( {\mathbf x} , \tilde{t} \leftarrow {\mathbf x}_{S} , \tilde{t}_{S} \right)
\, = \,
{\displaystyle
\frac{1}{4 \, \pi \, K_{0} \, \tilde{\tau}}}
\; \exp \left\{ - \,
{\displaystyle \frac{r^{2}}{4 \, K_{0} \, \tilde{\tau}}}
\right\} \;
\tilde{V}
\left( z , \tilde{t} \leftarrow z_{S} , \tilde{t}_{S} \right) \;\; ,
\label{propagator_reduced_lD_a}
\end{equation}
where the radial distance between the source at ${\mathbf x}_{S}$ and
the point ${\mathbf x}$ of observation is now defined as
\begin{equation}
r \, = \,
\left\{
\left( x - x_{S} \right)^{2} \, + \,
\left( y - y_{S} \right)^{2}
\right\}^{1/2} \;\; .
\end{equation}
Should the half--thickness $L$ be very large,
expression~(\ref{propagator_reduced_lD_a}) would boil down to the Gaussian
distribution~(\ref{propagator_reduced_3D_lD_a}) and the vertical propagator
$\tilde{V}$ would be given by the 1D solution $\mathcal{V}_{1D}$ of the
heat equation~(\ref{master_positron_3})

{\small
\begin{equation}
\tilde{V}
\left( z , \tilde{t} \leftarrow z_{S} , \tilde{t}_{S} \right)
\equiv
\mathcal{V}_{1D}
\left( z , \tilde{t} \leftarrow z_{S} , \tilde{t}_{S} \right) =
{\displaystyle \frac
{1}
{\sqrt{ 4 \, \pi \, K_{0} \, \tilde{\tau}}}} \;
\exp \left\{ - \, {\displaystyle
\frac{\left( z - z_{S} \right)^{2}}{4 \, K_{0} \, \tilde{\tau} \,}}
\right\} \;\; .
\label{propagator_reduced_1D}
\end{equation}}
But the diffusive halo has a finite vertical extent. We need to
implement the corresponding boundary conditions and impose that
the positron density vanishes at $z = \pm L$.

\noindent {\bf (i)}
A first approach relies on the method of the so--called  electrical
images and has been discussed in \cite{Baltz:1998xv}. Any point--like
source inside the slab is associated to the infinite series of its
multiple images through the boundaries at $z = \pm L$ which act as
mirrors. The n--th image is located at
\begin{equation}
z_{n} \, = \,
2 \, L \, n \; + \; \left( -1 \right)^{n} \, z_{S} \;\; ,
\end{equation}
and has a positive or negative contribution depending on whether $n$
is an even or odd number. When the diffusion time $\tilde{\tau}$ is
small, the 1D solution~(\ref{propagator_reduced_1D}) is a quite good
approximation. The relevant parameter is actually
\begin{equation}
\zeta \, = \,
{\displaystyle \frac{L^{2}}{4 \, K_{0} \, \tilde{\tau}}} \equiv
{\displaystyle \frac{L^{2}}{\lD^{2}}} \;\; ,
\label{definition_zeta}
\end{equation}
and in the regime where it is much larger than 1, the propagation
is insensitive to the vertical boundaries. On the contrary, when
$\zeta$ is much smaller than 1, a large number of images need to
be taken into account in the sum
\begin{equation}
\tilde{V}
\left( z , \tilde{t} \leftarrow z_{S} , \tilde{t}_{S} \right) \, = \,
{\displaystyle \sum_{n \, = \, - \infty}^{+ \infty}} \,
\left( -1 \right)^{n} \;
\mathcal{V}_{1D}
\left( z , \tilde{t} \leftarrow z_{n} , \tilde{t}_{S} \right) \;\; ,
\label{V_image}
\end{equation}
and convergence may be a problem.

\noindent {\bf (ii)}
It is fortunate that a quite different approach is possible in that
case. The 1D diffusion equation~(\ref{master_positron_3}) actually looks like
the Schr\"{o}dinger equation -- though in imaginary time -- that accounts for
the behaviour of a particle inside an infinitely deep 1D potential well
which extends from $z = - L$ to $z = + L$. The eigenfunctions of the
associated Hamiltonian are both even
\begin{equation}
\varphi_{n}(z) \, = \, \sin
\left\{ k_{n} \left( L - \left| z \right| \right) \right\}
\end{equation}
and odd
\begin{equation}
\varphi'_{n}(z) \, = \, \sin
\left\{ k'_{n} \left( L - z \right) \right\}
\end{equation}
functions of the vertical coordinate $z$. The wave--vectors $k_{n}$
and $k'_{n}$ are respectively defined as
\begin{equation}
k_{n} = \left( n - \frac{1}{2} \right)
{\displaystyle \frac{\pi}{L}} \;\; {\rm (even)}
\;\;\;\; {\rm and} \;\;\;\;
k'_{n} = n \, {\displaystyle \frac{\pi}{L}} \;\; {\rm (odd)} \;\; .
\end{equation}
The vertical propagator may be expanded as the series
{\footnotesize
\begin{equation}
\tilde{V}
\left( z , \tilde{t} \leftarrow z_{S} , \tilde{t}_{S} \right) =
{\displaystyle \sum_{n \, = \, 1}^{+ \infty}} \;\;
{\displaystyle \frac{1}{L}} \;
\left\{
e^{\displaystyle - \, \lambda_{n} \tilde{\tau}} \,
\varphi_{n} \left( z_{S} \right) \, \varphi_{n}(z)
\; + \;
e^{\displaystyle - \, \lambda'_{n} \tilde{\tau}} \,
\varphi'_{n} \left( z_{S} \right) \, \varphi'_{n}(z)
\right\} \;\; ,
\label{V_quantum}
\end{equation}}
where the time constants $\lambda_{n}$ and $\lambda'_{n}$ are
respectively equal to $K_{0} \, {k_{n}}^{2}$ and $K_{0} \, {k'_{n}}^{2}$.
In the regime where $\zeta$ is much smaller than 1, for very large
values of the diffusion time $\tilde{\tau}$, just a few eigenfunctions
need to be considered in order for the sum~(\ref{V_quantum}) to converge.
A close examination of these various expressions for $\tilde{G}$ indicate
that the energies $E$ and $E_{S}$ always come into play through the diffusion
length $\lD$ so that the positron propagator may be written as
\begin{equation}
G_{e^{+}} \!\! \left( {\mathbf x} , E \leftarrow {\mathbf x}_{S} , E_{S} \right) \, = \,
{\displaystyle \frac{\tau_{E}}{E_{0} \, \epsilon^{2}}} \;
\tilde{G} \left( {\mathbf x} \leftarrow {\mathbf x}_{S} ; \lD \right)
\;\; .
\label{positron_propagator_lD}
\end{equation}

%
%
\subsection{Secondary Positrons}
Like for antiprotons, a background of secondary positrons is produced by the spallation
of the interstellar medium by impinging high--energy particles. In that respect, the
Milky Way looks like a giant accelerator where cosmic rays play the role of the beam
whereas the galactic disc and its gas behave as the target.
%
%
The dominant mechanism is the collision of protons with hydrogen atoms at rest producing
charged pions $\pi^{\pm}$ which decay into muons $\mu^{\pm}$. The latter are also unstable
and eventually lead to electrons and positrons through the chain
%
\begin{flushright}
\begin{tabular}{c c c c c r}
${\rm p \; + \; H} \; \longrightarrow \; {\rm X} \;\; +$ &
$\!\!\!\!\!\!\! \pi^{\pm} \!\!\!\!\!\!\!$ & & & &
\hspace{1.0cm}
\label{pp_pion_chain}
\\
& $\!\!\!\!\! \pi^{\pm} \!\!\!\!\!$ & $\! \longrightarrow \; \nu_{\mu} \; +$
& $\!\!\!\!\!\!\! \mu^{\pm} \!\!\!\!\!\!\!$ & & \\
& & & $\!\!\!\!\!\!\! \mu^{\pm} \!\!\!\!\!\!\!$ &
$\!\longrightarrow \; \nu_{\mu} \; + \; \nu_{e} \; + \; e^{\pm} \;\; .$ & \\
\end{tabular}
\end{flushright}
%
Below $\sim$ 3 GeV, one of the protons is predominantly excited to a $\Delta$
resonance which subsequently decays into either a neutral or a charged pion.
The former species produces gamma rays whereas the latter particle decays into
positrons.
Above $\sim$ 7 GeV, pion production is well described
in the framework of the scaling model. Various parameterizations are given in the
literature~\cite{badhwar_1977,Tan:1984ha} for the Lorentz invariant (LI) cross
section $E_{\pi} \, {d^{3} \sigma}/{d^{3}p_{\pi}}$.
Positrons may also be produced through kaons although this channel is rare.
The positron production cross sections of these processes have been carefully
computed in the appendices of~\cite{1998ApJ...493..694M}.
Notice also that useful parametric expressions for the yield and spectra of the
stable secondary species produced in p--p collisions have been derived from
experimental data and summarized in \cite{Kamae:2006bf}.
%
\begin{figure}[h!]
\begin{center}
\noindent
\includegraphics[width=0.75\textwidth]{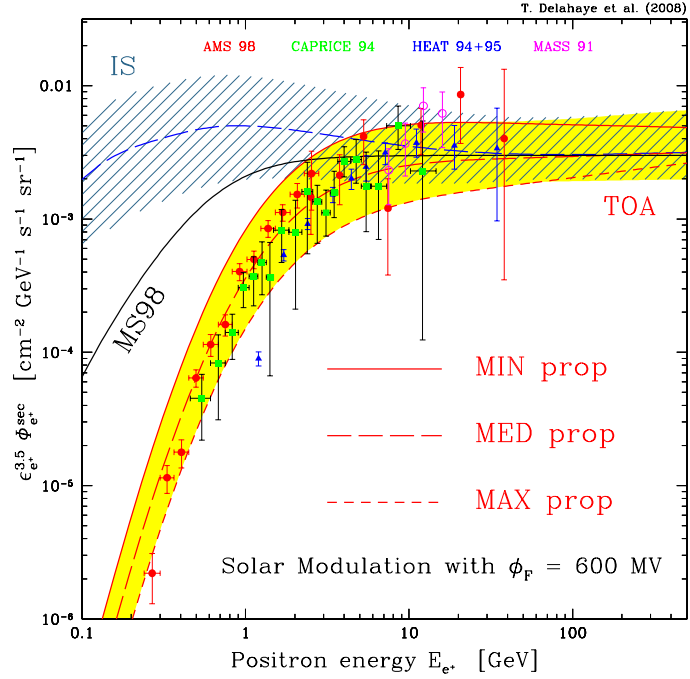}
\end{center}
\vskip -0.25cm
\caption{
Secondary positron flux as a function of the positron energy.
The blue hatched band corresponds to the cosmic ray propagation
uncertainty on the interstellar prediction whereas the yellow strip
refers to the top--of--atmosphere fluxes.
The long--dashed curves feature our reference model MED with
a differential production cross section borrowed from~\cite{Kamae:2006bf}
and the most recent measurements by BESS~\cite{2007APh....28..154S}
of the cosmic ray proton and helium fluxes.
The MIN, MED and MAX propagation parameters are displayed in
table~\ref{tab_prop}.
The observations by
CAPRICE~\cite{2000ApJ...532..653B} (green squares),
HEAT~\cite{1997ApJ...482L.191B} (blue triangles),
MASS~\cite{2002A&A...392..287G} (open circles)
and AMS~\cite{Alcaraz:2000bf,Aguilar:2007yf} (red dots)
are also indicated.
}
\label{fig:secondary_positrons}
\end{figure}
%

%
%
Cosmic ray protons with energy $E_{\rm p}$ induce a production of positrons
per hydrogen atom with a rate
\beq
d{\Gamma}_{e^{+}}^{\rm sec}(E_{e}) \, = \,
{\displaystyle \frac{d\sigma}{dE_{e}}}(E_{\rm p} \to E_{e})
\times \beta_{\rm p} \times
\left\{ dn_{\rm p} \equiv \psi_{\rm p}(E_{\rm p}) \times dE_{\rm p} \right\} \;\; .
\eeq
This leads to the source term
\beq
q_{e^{+}}^{\rm sec} \!\! \left( {\mathbf x} , E_{e} \, \right) \, = \,
4 \; \pi \; n_{H}({\mathbf x}) \,
{\displaystyle \int} \, \Phi_{\rm p} \! \left( {\mathbf x} , E_{\rm p} \, \right)
\times dE_{\rm p} \times
{\displaystyle \frac{d\sigma}{dE_{e}}}(E_{\rm p} \to E_{e}) \;\; .
\label{source_sec_pos}
\eeq
That relation can be generalized in order to incorporate cosmic ray helium
nuclei as well as interstellar helium.
The gas of the galactic plane is generally assumed to be homogeneously
spread. Because positrons detected at the Earth originate mostly from the
solar neighborhood, we can safely disregard the space dependance of the
proton and helium fluxes. Making use then of the
measurements~\cite{2001ApJ...563..172D,2007APh....28..154S} in
relation~(\ref{source_sec_pos}), we can express the secondary positron flux as
\beq
\Phi_{e^{+}}^{\rm sec} \!\! \left( \odot , \epsilon \equiv {E_{e}}/{E_{0}} \right)
\, = \, {\displaystyle \frac{\beta_{e^{+}}}{4 \, \pi}} \times
{\displaystyle \frac{\tau_{E}}{\epsilon^{2}}} \times
{\displaystyle \int_{\epsilon}^{+ \infty}} d\epsilon_{S} \times
\tilde{I} \left( \lD \right) \times
q_{e^{+}}^{\rm sec} \!\! \left( \odot , \epsilon_{S} \, \right) \;\; .
\label{pos_flux_simple}
\eeq
The integral $\tilde{I}$ is the convolution of the reduced positron Green
function $\tilde{G}$ over the galactic disc alone
\beq
\tilde{I} \left( \lD \right) \, = \,
{\displaystyle \int}_{\rm \!\! disc} \!\! d^{3}{\mathbf x}_{S} \;\;
\tilde{G} \left( {\mathbf x}_{\odot} \leftarrow {\mathbf x}_{S} ; \lD \right) \;\; ,
\label{I_DH_b}
\eeq
and depends only on $\lD$ as discussed above.

%
%
This method has been recently used~\cite{eplus_sec} to derive the positron flux
featured in figure~\ref{fig:secondary_positrons}.
At 1 GeV, the width of the IS uncertainty strip corresponds to an increase by
a factor of $\sim$ 6 between the smallest and the largest positron fluxes allowed
by the B/C constraint. That factor decreases down to 3.9 at 10 GeV and reaches
2.9 at 100 GeV.
Once modulated with a Fisk potential $\phi_{\rm \, F}$ of 600 MV, the blue hatched
region is transformed into the yellow TOA band. Quite surprisingly, the MIN model
(red solid curve) corresponds now to the {\bf maximal} secondary positron flux whereas
the MAX configuration (red short--dashed line) yields the {\bf minimal} prediction.
At fixed energies $\epsilon$ and $\epsilon_{S}$, the smaller the diffusion coefficient
$K_{0}$, the smaller the diffusion length $\lD$ and the larger the integral~(\ref{I_DH_b}).
The latter reaches its maximal value of 1 whenever $\lD$ is smaller than the disc
half--thickness $h$ of 100~pc.

%
%
\subsection{DM signals in Cosmic Positrons}
An excess of the positron measurements with respect to the astrophysical background
seems to appear above 10 GeV. This trend was present in the HEAT
data~\cite{1997ApJ...482L.191B} and has been recently nicely confirmed by the
PAMELA observations~\cite{Adriani:2008zr} of the positron fraction
\beq
{\displaystyle \frac{e^{+}}{e^{+} \, + \, e^{-}}} \equiv
{\displaystyle \frac{\Phi_{e^{+}}^{\rm \, tot}}
{\Phi_{e^{+}}^{\rm \, tot} \, + \, \Phi_{e^{-}}^{\rm \, tot}}} \;\; .
\label{positron_fraction}
\eeq
Different astrophysical contributions to the positron fraction in the 10~GeV region
have already been been explored in~\cite{1997ApJ...482L.191B}. More accurate and
energy extended data will
shed light on the effective presence of a bump in the positron absolute flux and on
its physical interpretation.
A possible explanation relies on DM species annihilating
in the galactic halo \cite{Baltz:1998xv,Hooper:2004bq}.
This hypothesis has been strongly revived~\cite{Cirelli:2008pk,Bergstrom:2008gr}
with the PAMELA observation of what seems a clear positron excess above 10 GeV.
%
\begin{figure}[h!]
\begin{center}
\vskip -1.5cm
\noindent
\includegraphics[angle=270,width=1.05\textwidth]{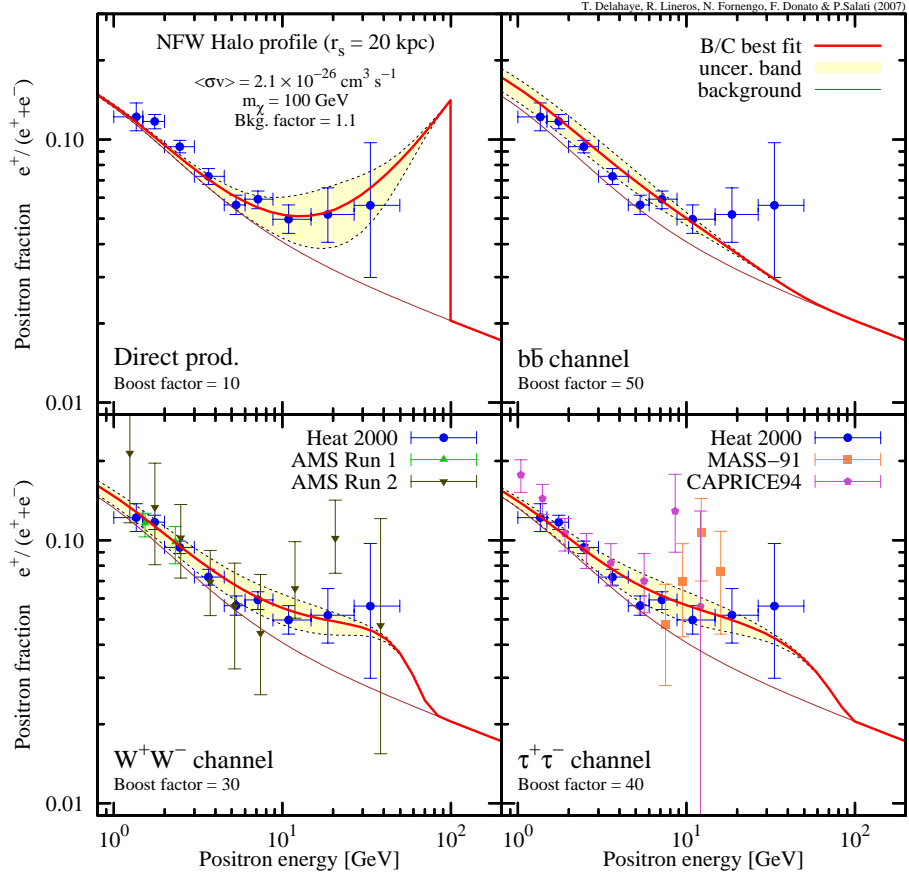}
\end{center}
\vskip -0.75cm
\caption{
The positron fraction ${e^{+}}/{(e^{-} + e^{+})}$ is plotted versus the positron
energy $E$ for a 100 GeV WIMP in the case of a NFW DM profile.
The four panels refer to different annihilation final states~: direct $e^{+} e^{-}$ production
(top left), $b \bar{b}$ (top right), $W^{+} W^{-}$ (bottom left) and $\tau^{+} \tau^{-}$
(bottom right).
In each panel, the brown thin solid line stands for the positron background borrowed
from~\cite{1998ApJ...493..694M} and parameterized by~\cite{Baltz:1998xv}.
The red thick solid curve refers to the {\bf total} positron flux where the signal
is propagated with the best--fit choice of the astrophysical parameters, \ie
the configuration MED of table~\ref{tab_prop}.
The yellow area features the total uncertainty band arising from cosmic ray propagation.
The different models found in~\cite{2001ApJ...555..585M} to be compatible with the B/C
ratio all yield a positron fraction which is enclosed inside this yellow strip.
Experimental data from
HEAT~\cite{1997ApJ...482L.191B},
AMS~\cite{Alcaraz:2000bf,Aguilar:2007yf},
CAPRICE~\cite{2000ApJ...532..653B} and
MASS~\cite{2002A&A...392..287G} are also presented for comparison.
Figure from \cite{2008PhRvD..77f3527D}.
}
\label{fig:pf_CR_band_100_GeV}
\end{figure}
%

%
%
Such an interpretation, though very exciting, is at some point limited by the uncertainties
in the halo structure and in the cosmic ray propagation modeling.
Equations~(\ref{psi_convolution_positron}) and (\ref{positron_propagator_lD}) can be
combined to yield the positron flux generated by the WIMP annihilations taking place
within the Milky Way diffusive halo
\beq
\Phi_{e^{+}}^{\rm DM} \!\! \left( \odot , \epsilon \equiv {E_{e}}/{E_{0}} \right) \, = \,
{\mathcal F} \times {\displaystyle \frac{\tau_{E}}{\epsilon^{2}}} \times
{\displaystyle \int_{\epsilon}^{m_{\chi}/E_{0}}} d\epsilon_{S} \; g(\epsilon_{S}) \;
\tilde{I}_{\rm DM} \left( \lD \right) \;\; .
\label{flux_positron_DM}
\eeq
The information related to particle physics has been factored out in
\beq
{\mathcal F} \, = \,
{\displaystyle \frac{\beta}{4 \pi}} \; \xi^{2} \;
{\left\langle \sigma_\mathrm{ann} v \right\rangle} \,
\left\{ {\displaystyle \frac{\rho_{\odot}}{m_{\chi}}} \right\}^{2} \;\; .
\label{factor_F}
\eeq
The energy distribution $g(\epsilon_{S})$ describes the positron spectrum
at the source and depends on the details of the WIMP annihilation mechanism.
The halo integral $\tilde{I}_{\rm DM}$ is the convolution of the reduced positron
propagator $\tilde{G}$ with the square of the DM galactic density
\beq
\tilde{I}_{\rm DM} \left( \lD \right) \, = \,
{\displaystyle \int}_{\rm \!\! DH} \!\! d^{3}{\mathbf x}_{S} \;
\tilde{G} \left( {\mathbf x}_{\odot} \leftarrow {\mathbf x}_{S} ; \lD \right) \;
\left\{ {\displaystyle \frac{\rho_{\chi}({\mathbf x}_{S})}{\rho_{\odot}}} \right\}^{2}
\;\; .
\label{I_DH_c}
\eeq
%
%
In figure~\ref{fig:pf_CR_band_100_GeV}, the positron fraction~(\ref{positron_fraction})
is presented as a function of the positron energy $E$. The total positron flux at the Earth
\beq
\Phi_{e^{+}}^{\rm \, tot} \, = \,
\Phi_{e^{+}}^{\rm DM} \, + \, \Phi_{e^{+}}^{\rm sec}
\eeq
encompasses the annihilation signal and a background component for which the results
of~\cite{1998ApJ...493..694M} as parameterized by~\cite{Baltz:1998xv} have been used --
see the brown thin solid lines. The mass of the DM species is 100~GeV and a NFW profile
has been assumed. The observations featured in the various panels are indications of a
possible excess of the positron fraction for energies above 10~GeV. Those measurements
may be compared to the red thick solid curves that correspond to the MED configuration
of table~\ref{tab_prop}. In order to get a reasonable agreement between the DM predictions
and the data, the annihilation signal has been boosted by an energy--independent factor
ranging from 10 to 50 as indicated in each panel.
%
%
As is clear in the upper left panel, the case of direct production offers a very good
agreement with the positron excess. Notice how well all the data points lie within
the yellow uncertainty band. A boost factor of 10 is enough to obtain an excellent agreement
between the measurements and the median flux. A smaller value would be required for a flux
at the upper envelope of the uncertainty strip. The $W^{+} W^{-}$ and $\tau^{+} \tau^{-}$
channels may also reproduce reasonably well the observations, especially once the uncertainty
arising from cosmic ray propagation throughout the Milky Way diffusive halo is taken into
account. They need though larger boost factors of the order of 30 to 40. On the contrary,
softer production channels, like the $b \bar{b}$ case, are unable to match the features of
the positron bump.

%
%
\section{Conclusions}
Crucial information on the astronomical dark matter pervading the Galaxy
will be provided by crossing the positron~\cite{Adriani:2008zr} and
antiproton~\cite{Adriani:2008zq} observations. An excess seems to be
present in the positron flux but does not show up in the antiproton
signal. This may lead to strong constraints on the properties of
the DM species as mentioned for instance in~\cite{Cirelli:2008pk,Donato:2008jk}.
Pure leptonic models are a priori favored since even a small branching ratio
to quarks or gauge bosons would translate into an overproduction of antiprotons
which is not seen. Some attention needs to be paid though insofar as antiprotons
could be rehabilitated depending on how cosmic ray transport is modeled.
The forthcoming measurements will be crucial to ascertain the presence of
WIMPs in outer space and to study their nature.

%
%

%
\bibliographystyle{plainyr}
\bibliography{book}

\end{document}